%% file: main.tex
\documentclass[twocolumn]{aastex701}
\newcommand{\eiso}{$E_{\rm \gamma,iso}$}

\newcommand{\ltev}{$L_{\rm TeV,11h}$}
\newcommand{\lx}{$L_{\rm X,11h}$}
\newcommand{\gray}{$\gamma$-ray}

\newcommand{\lt}{\ensuremath <}
\newcommand{\gt}{\ensuremath >}

\usepackage{amssymb}
\usepackage{amsmath}
\usepackage{lipsum}
\usepackage{appendix}
\usepackage{xcolor}

\pdfoutput=1 

\usepackage{hyperref}

\usepackage[T1]{fontenc}

\usepackage{graphicx}
\usepackage{subcaption}

\usepackage{amssymb}
\usepackage{lipsum}

\usepackage{siunitx}

\usepackage{placeins}

\begin{document}

\title{Chasing Gamma-Ray Signals from Binary Neutron Star Coalescences with the Cherenkov Telescope Array: Prospects and Observing Strategies}

\input{cta-authorlist}

\begin{abstract}

The detection of gravitational waves (GWs) from a binary neutron star (BNS) merger by Advanced LIGO and Advanced Virgo (GW170817),
together with its electromagnetic counterpart, the short gamma-ray burst GRB~170817A, heralded the birth of multi-messenger astronomy. 
The detection of TeV emission from GRBs motivates follow-up observations with the Cherenkov Telescope Array Observatory (CTAO), ideal for detecting such signals due to its unprecedented sensitivity, rapid response, and wide-field survey capabilities.
The aim of this work is to evaluate GeV--TeV GW follow-up strategies for CTAO using a multi-step simulation pipeline and to estimate the expected rate of joint GW-GRB detections during observing run O5.

Using a simulated sample of BNS systems with corresponding GW detections, gamma-ray emission is simulated through phenomenological prescriptions based on the observed population of short GRBs, including off-axis jet scenarios. CTAO observations are simulated to account for instrument response, sky tiling strategies, integration times, and varying observing conditions. Strategies with variable and constant integration times are investigated.

We find that, via an optimized follow-up strategy, about 5\% of simulated GW-associated short GRBs produce GeV--TeV radiation detectable by CTAO. 
Detectability is strongly influenced by the jet opening angle and viewing angle, suggesting that even rough estimates of the viewing angle in GW alerts could enhance targeting. This framework motivates future follow-ups of GW-detectable events, including neutron star-black hole mergers, and further supports the development of advanced strategies incorporating galaxy distributions and synergies with future detectors such as the Einstein Telescope.

\end{abstract}

\section{Introduction}
\label{Intro}

On August 17, 2017 a gravitational wave (GW) signal from the inspiral of a binary neutron star (BNS) merger was observed for the first time \citep{2017PhRvL.119p1101A} by Advanced LIGO \citep{2015CQGra..32g4001L} and Advanced Virgo \citep{2015CQGra..32b4001A}; $\sim$2 seconds after this GW event, Fermi \citep{2017ApJ...848L..14G} and INTEGRAL \citep{2017ApJ...848L..15S} detected a short Gamma-Ray Burst (GRB) in GRB 170817A. The association of GW170817 and GRB 170817A represents the first direct evidence that BNS mergers can produce short GRBs (sGRB), confirmed by the VLBI observations of a successful, structured jet \cite{ghirlanda_compact_2019}.  Long GRBs produce gamma-ray photons reaching TeV energies, as proven by the recent observations of GRB 190114C, GRB 201216C, and GRB 201015A by MAGIC \citep{2019Natur.575..455M,MAGIC_201216C,2020GCN.28659....1B},  GRB 180720B and GRB 190829A by H.E.S.S. \cite{2019Natur.575..464A,2021Sci...372.1081H} and GRB 221009A by LHAASO \citep{LHAASO_221009A,LHAASO_221009A_secondpaper}. Delayed by bright moonlight until 1.33 days post-event, observations of GRB 221009A using LST-1, the first Large-Sized Telescope of the upcoming Cherenkov Telescope Array Observatory (CTAO), reached a significance of $4.1\sigma$, resulting in the derivation of upper limits \citep{abe2025grb}. Lastly, MAGIC observed the short GRB~160821B, obtaining a hint of TeV emission at a 3$\sigma$ level \citep{2021ApJ...908...90A}. 
Motivated by these results, the scope of this work is to determine not only the prospects of joint GW-GRB detections with CTAO, but also how the choice of observational strategy affects these prospects. 

\subsection{The Cherenkov Telescope Array Observatory}
CTAO is poised to be the next-generation of Imaging Atmospheric Cherenkov Telescopes (IACTs), covering a range in the very-high-energy (VHE) gamma-ray regime between 20~GeV and 300~TeV (see \citealp{2013APh....43....3A}). So far, no VHE emission coincident with a GW event has been observed in the VHE band by any IACT. Upper limits were derived in the case of GW170817, both several hours \citep{2017ApJ...850L..22A} and months post-merger \citep{2020ApJ...894L..16A,2022icrc.confE.944S}. Instead, in the case of GRB 211211A, a joint detection of kilonova emission was reported at a distance of $\sim$ 350 Mpc, closely resembling AT2017g, the kilonova counterpart to GW170817, together with GeV emission by Fermi-LAT \citep{rastinejad2022kilonova}. During this period, the gravitational-wave interferometer network was undergoing commissioning towards the observing run O4, and consequently no gravitational-wave detection was possible.

CTAO will provide an order of magnitude improvement in sensitivity compared to the current generation of IACTs, swift slewing capabilities reaching 180$^{\circ}/20s$, and a large field-of-view (FoV) between 4-10$^{\circ}$ depending on the telescope: these characteristics make it an ideal instrument to search for VHE emission from short GRBs associated to GW events. 

\subsection{LIGO-Virgo-KAGRA and IACTs}
The fourth LIGO-Virgo-KAGRA (LVK) observing run started on May 24, 2023 and finished on November 18, 2025\footnote{\url{https://observing.docs.ligo.org/plan/}}. Upgrades of Advanced LIGO, Advanced Virgo and KAGRA \cite{2019CQGra..36p5008A} interferometers will improve the broadband sensitivity towards the fifth LVK observing run, O5. O5 is currently planned for the beginning of 2028, a timeline consistent with the schedule of CTAO. The increased sensitivity will allow us to explore a much wider volume of the Universe, with the consequent increase in the GW detection rate  \cite{2022ApJ...924...54P} and potentially in the multi-messenger detection rate with respect to the previous observing runs. 

Yet, even with improved sensitivity in current and next-generation gravitational-wave interferometers, significant localization uncertainties will persist for a large fraction of detections, largely depending on the size of the detector network. Improvements throughout the years on observation strategies (e.g. \citealp{singer2025optimal}) and observation campaign coordination tools (e.g. \citealp{wyatt2020gravitational}) are key to maximize the detection odds of the counterparts of gravitational wave detections, as those from compact binary coalescences. These automatic handling and scheduling strategies are crucial in small and medium-sized FoV telescopes, as in the case of the IACT, and have been optimized to cover rapidly large regions in the sky, as shown in \citep{ashkar2021hess}. In addition, recent GW follow-ups by IACTs have been shown to have the potential to cover a large portion of the uncertainty region of GW, as shown by the combined LST and MAGIC observations of S241125n \citep{2024GCN.38443....1P}, putting upper limits on the gamma-ray emission above 300 GeV, proving the capabilities of TeV observations to set useful constraints to the emission models.   \citep{2025_S241125n_MAGIC-LST_inPrep}. 

\subsection{GW Follow-up Campaigns}
The capability of CTAO to perform the EM follow-up of GW events, as well as possible observational strategies, has already been studied in a few works (\citealp{bartos_strategies_2018}, \citealp{2018JCAP...05..056P}).
In particular, \cite{2018JCAP...05..056P} investigated the prospects for joint GW and VHE EM observations with Advanced Virgo, Advanced LIGO, and CTAO, based on detailed simulations of BNS merger accompanied by short GRBs, with a focus on \textit{on-axis} sources. These refer to GRBs observed within the opening angle of the jet, as opposed to \textit{off-axis}, which are those observed from outside the opening angle of the jet. They proposed an optimized observational strategy for CTAO EM follow-up of GW events, involving increasing exposure times for consecutive observations, and demonstrated that this approach enhances the likelihood of detecting VHE EM counterparts. 
Other works (\citealp{2014MNRAS.443..738B}, \citealp{bartos_gravitational-wave_2019}, \citealp{banerjee_pre-merger_2023}, \citealp{mondal_follow-up_2024}) explored the capability of CTAO to follow-up GW events and to detected the electromagnetic counterparts in the VHE domain. In \citealp{2014MNRAS.443..738B}, the authors focused on poorly localized sources, demonstrating that CTAO will be able to detect short GRBs associated with GW events even if it needs to survey a sky area of $\sim$ 1000 deg$^2$ and if the observations are delayed by $\sim$ 100 s following the onset of gamma-ray emission.  The prospects for CTAO with the next generation of GW interferometers such as the Einstein Telescope\footnote{\url{https://www.et-gw.eu/}} have been briefly explored (\citealp{banerjee_pre-merger_2023}, \citealp{colombo_multi-messenger_2025}). Interestingly, \citep{banerjee_pre-merger_2023} showed that early-warning GW alerts—i.e., alerts issued prior to the merger of a binary system—could enable the detection of early VHE emission with CTAO. Conversely, other studies \citep{pellouin_very_2024, hope_tev_2025} have attempted to model the high-energy gamma-ray components of the GRB associated with GW170817, reaching differing conclusions regarding the detectability of VHE emission from the associated off-axis and on-axis GRBs. \cite{yuan2022gev} examined the scenario of compact-object mergers embedded within the disk of Active Galactic Nucleus (AGN), where the predominant contribution to the gamma-ray emission would originate from external inverse Compton scattering of isotropic thermal photons from the disk. Their prospects show that for long-lasting jets of $T_{dur} \sim 10^2-10^3 s$ CTAO will be able to detect 25-100 GeV emission out to a redshift $z =1$.  Recent work has explored the potential of detecting EM signals from stochastic GW backgrounds using a time-domain cross-correlation approach \citep{kuralkar2025new}, which is complementary to the approach described in this work, that focuses in matched-filtering-detected compact binary coalescences (CBCs) with two neutron stars as compact objects.

\subsection{VHE Emission of Short GRBs}
Short and long GRBs are expected to have different progenitors: long GRBs are associated with core-collapse of massive stars, while short GRBs are associated with the merger of compact binaries containing at least a NS (\citealp{2014ARA&A..52...43B}; see, however, \citealp{2022Natur.612..223R,2024Natur.626..737L}).

The existence of VHE emission—extending beyond 10 TeV—has been observed in long GRBs, i.e., from on-axis beamed jets produced after the collapse of the central object. Given the possibility of a similar jet structure in both long and short GRBs \citep{ghirlanda_short_2011}, TeV emission may be a common feature in short GRBs, yet this remains to be confirmed. The prevailing physical explanation for this emission is that VHE photons are produced via the synchrotron self-Compton (SSC) process at the shock formed in the afterglow phase by the interaction of the relativistic jet with the interstellar medium  (see e.g. \citep{190114C_second_MAGIC_paper}, and \citep{Miceli_Nava_review} for a more extended list of references). 
The detectability of VHE photons from GW counterparts depends on several poorly constrained factors, including the microphysics driving jet processes, the geometry of the jetted emission (e.g., jet opening angle and structure), and the properties of the surrounding environment, such as the interstellar medium (ISM) density and distribution. 

The EM counterparts of GW events are believed to be produced in both on-axis and off-axis GRBs, when structured jets are assumed \citep{2018MNRAS.473L.121K}, as confirmed by very-long-baseline interferometry observations of GW170817 \citep{ghirlanda_compact_2019}. The detailed structure of the jet affects off-axis observations by introducing a delay in the peak of the afterglow emission and causing an energy-dependent shift in the prompt emission (\citealp{2024MNRAS.tmp..115B},\citealp{salafia_structure_2022}). These effects also depend on the jet opening angle, for which only limited constraints are currently available \citep{2023ApJ...959...13R, 2015ApJ...815..102F}.\\
The VHE emission is sensitive to these parameters, and constraining it provides unique post-merger signatures that are inaccessible in other EM bands. This work further investigates their impact on the detectability of the jetted emission from GW events with CTAO.

\subsection{This Work}
Compared to previous work, we implemented a complete simulation chain, starting from GW events originating from BNS mergers and detectable by the LVK instruments during the O5 observing run. The simulation includes physically motivated and phenomenologically supported multi-band and VHE emission from a structured jet in the afterglow phase, as well as a full modeling of the CTAO instrument response to the expected time-variable spectral energy distributions as described in the following sections. The events are further processed through a realistic scheduling framework that takes into account their visibility from both the southern and northern CTAO arrays in operational darkness conditions, along with an appropriate observing strategy. These consist of multiple consecutive observations, called tiling, which are strategically selected to provide coverage of a large fraction of the GW uncertainty region.

The paper is organized as follows. In Section \ref{sec:GW} we describe the sample of simulated BNS systems and the associated GW signals that we use in this work. In Section \ref{sec:VHE_emission} we detail how we simulated the VHE emission associated with BNS mergers. In Section \ref{sec:CTAO} we describe the characteristics of CTAO. In Sections \ref{sec:CTAsim} and \ref{sec:obs-strategy} we present the simulations of the CTAO follow-up of BNS mergers and propose observational strategies to optimize the probability of detection of EM counterparts to GW events. Finally, in Section \ref{sec:discussion} we describe and discuss our results, and in Section \ref{sec:conclusion} we present our conclusions.

\section{Simulated GW events from BNS mergers}\label{sec:GW}

The realistic ensemble of simulated BNS mergers and the associated GW signals presented in \citet{2022ApJ...924...54P}, that has been released publicly on Zenodo \citep{2021zndo...5206853S}, is used in these simulations. This BNS sample, consisting of 2307 events, defines the size of our full GW-GRB simulation set, as each BNS has been connected to a GRB emission (see Sec. \ref{sec:VHE_emission}.) The sample has been built considering realistic astrophysical distributions of masses, spins, distances, and sky locations. Specifically, the NS component masses were randomly extracted from a normal distribution with mean 1.33 M$_\odot$ and standard deviation 0.09 M$_\odot$, consistent with that inferred from measurements related to binary systems
in the Galaxy \citep{2016ARA&A..54..401O}. The NS spins are aligned or anti-aligned,  with uniformly distributed magnitudes smaller than 0.05: the maximum allowed value is consistent with the spin of the most rapidly rotating
pulsar found in a binary system, i.e. PSR J0737-3039A  \citep{2003Natur.426..531B}. The position and the orientation of the binaries are distributed isotropically, and the  redshifts are drawn uniformly in co-moving rate density, employing cosmological parameters from \citet{2016A&A...594A..13P}. For each BNS merger, the expected GW inspiral signal has been simulated and then convolved with the GW detector responses, considering different GW detector network configurations. For this work, we select the simulations performed assuming a network composed by Advanced LIGO, Advanced Virgo and KAGRA with the sensitivities expected to be reached in the fifth LIGO-Virgo-KAGRA observing run (O5), whose timeline is consistent with the one of CTAO; these sensitivities have been released in \citet{LIGO-T2000012}.  
The obtained BNS simulations have then been analyzed with the matched filtering technique \citep{wainstein63,2014PhRvD..90h2004D,2015PhRvD..91d2003V,2016CQGra..33q5012A,2016CQGra..33u5004U,2012ApJ...748..136C,2017PhRvD..95d2001M,2017ApJ...849..118N}, assuming a 70\% independent duty cycle for each interferometer. The signals are considered as GW candidates if they are detected with a network signal-to-noise ratio above 8, even if observed with a single interferometer: this approach has been shown to be representative of the public GW alerts sent during the third LIGO-Virgo-KAGRA observing run (O3) \citep{2022ApJ...924...54P}. Finally, for each GW simulated candidate the associated sky localization has been estimated with BAYESTAR, that is a rapid Bayesian position reconstruction code that computes source location using the output from the detection pipelines \citep{2014ApJ...795..105S}. Included in this localization information is the $90$\% credible region (C.R.) skymap, utilized in all observation strategies and presented in \ref{sec:obs-strategy}.

\section{Estimation of the VHE emission from phenomenological prescriptions}\label{sec:VHE_emission}
The estimate of the VHE emission is performed adopting a simple, phenomenological model. All BNS merger events included in the simulated catalog (see Sec.~\ref{sec:GW}) are assumed to successfully launch a relativistic collimated outflow (i.e. jet), whose energy dissipation leads to emission of radiation known as short GRB. This hypothesis is adopted for simplicity, as considerations of  the uncertainties in jet launching mechanism and propagation lie beyond the scope of this work.
Yet, only a fraction of events may be capable of launching a jet that successfully penetrates the ejecta and achieves a breakout \citep{nikhil_sgrb_2022,salafia_incidence_2022,colombo_multimessenger_2022A}, which is subject to discussion in Sec.~\ref{sec:discussion}. 

Based on the information available for each simulated GW event (and in particular distance and viewing angle $\theta_{\rm view}$, defined as the angle between the jet axis and the line of sight to the observer), the jet-related emission in the CTAO energy range is estimated from phenomenological prescriptions based on the currently available observations of short GRBs at different wavelengths and on the current knowledge of VHE emission from GRBs. The adopted approach does not require the specific population of particles or the specific radiative mechanism responsible for the production of \gray s. Since we are interested in long-lasting radiation, we focus on VHE emission produced in the context of the interactions between the jet and the surrounding medium, i.e., afterglow radiation.

The information currently available from GRBs at VHE includes: i) five detections of long GRBs either by MAGIC \citep{2019Natur.575..455M,MAGIC_201216C}, H.E.S.S. \citep{2019Natur.575..464A,2021Sci...372.1081H}, or LHAASO \citep{LHAASO_221009A}, ii) one hint of emission from a short GRB by MAGIC \citep{2021ApJ...908...90A}, and iii) flux upper limits inferred from observations of a large number of (mostly long) GRBs by MAGIC,  H.E.S.S., and HAWC \citep{HESS_UL,MAGIC_UL,HAWC_UL}. 

Past detections of long GRBs between $0.1$ and $10$\,TeV have shown that the luminosity emitted in this energy range is comparable to the luminosity emitted in the soft X-ray band \citep{190114C_second_MAGIC_paper,Nava_review,Miceli_Nava_review}, and decays in time at a similar rate \citep{2021Sci...372.1081H}. 
Given a current lack of observational evidence of short GRBs at TeV energies, we assume here that TeV radiation in short GRBs behaves in the same way, i.e. that its TeV luminosity is comparable to the X-ray one and decays at a similar rate. If this assumption would not be proved by future observations, part of the results in this paper might require substantial revision.

Spectra of the VHE detected GRBs are consistent with a photon index around $\alpha=-2$ (in the notation $dN/dE\propto E^{\alpha}$, where $dN/dE$ is the number of photons per unit of energy), ranging from $-1.6$ (\citealp{2019Natur.575..464A}, although poorly constrained) to $-2.5$ or even softer, depending on time and the Extragalactic Background Light (EBL) model \citep{2019Natur.575..455M,LHAASO_221009A_secondpaper}. Since the sample of detected GRBs is still very limited, interesting information can come also from observed GRBs with no evidence of VHE excess. In these cases flux upper limits can be estimated and compared with the properties of VHE detected GRBs. 

Starting from these pieces of information, we predict VHE afterglow light-curves and spectra for an observer located at an angle $\theta_{\rm view}$ from the jet axis. This requires to specify a jet structure, i. e., a description of how the kinetic energy density and Lorentz factor of the jet depend on the angle $\theta$ from the jet axis.

In the next sections, we specify all the assumptions on the adopted parameters  and give the details of the method that has been applied to simulate light-curves and spectra of VHE radiation.

\subsection{Jet properties and structure}
We consider a jet with a Gaussian structure in energy and bulk Lorentz factor \citep{zhang2002,kumar2003}:

\begin{eqnarray}
    \epsilon_{\rm k}(\theta) = \epsilon_{\rm k,core}\,e^{\left(-\frac{\theta^2}{\theta_{core}^2}\right)}\\
    \Gamma_0(\theta) = (\Gamma_{0,core}-1) \, e^{\left(-\frac{\theta^2}{2\theta_{core}^2}\right)} + 1
\end{eqnarray}

Here $\theta$ is the angle from the jet axis and ranges from $\theta=0^\circ$ and $\theta=\theta_{\rm MAX}$, where $\theta_{\rm MAX}$ is defined as the minimum between 90$^\circ$ and the angle that verifies the condition $\Gamma_0(\theta_{\rm MAX})=5$. This condition is introduced to limit the computation time (see similar approaches in \citep{pellouin2024,ryan2020}).
The energy per unit solid angle of the core is $\epsilon_{\rm k,core}\sim E_{\rm k,core}/\pi\theta_{\rm core}^2$.  
Hence, the jet structure depends on three quantities: the opening angle $\theta_{\rm core}$, the initial Lorentz factor $\Gamma_{\rm 0,core}$, and the core kinetic energy $E_{\rm k,core}$. The value of $\theta_{\rm core}$ is randomly assigned according to the distribution based on the few estimates inferred from the population of short GRBs seen on-axis. 
We use a lognormal distribution with $\langle Log(\theta_{\rm core}/\rm deg)\rangle= 1.15$ (corresponding to 14$^\circ$) and $\sigma_{Log\theta_{core}} = 0.2$ \citep{fong15}. 
Lacking a statistical study of bulk Lorentz factors in short GRB jets, its value has been randomly assigned from a lognormal distribution peaked at 
$\Gamma_{\rm 0,core}=200$ (a value similar to the bulk Lorentz factor adopted to model emission from GRB~170817A \citep{ghirlanda_compact_2019,salafia2019}) and standard deviation $\sigma=0.2$.

The kinetic energy of the jet $E_{\rm k,core}$ at the beginning of the afterglow emission is derived from the $\gamma$-ray energy emitted during the prompt phase by assuming an efficiency $\eta_\gamma$ for the prompt emission: $E_{\rm k,core} \sim E_{\rm \gamma}\,(1-\eta_\gamma)/\eta_\gamma$. Note that the initial jet energy can be computed as $E_{\rm 0,jet} = E_{\gamma} / \eta_{\gamma}$.
The collimation corrected energy $E_\gamma$ is related to the more commonly used isotropic equivalent energy \eiso: $E_{\rm \gamma} \sim E_{\rm \gamma,iso}\,(1-\cos{\theta_{\rm core}})/2$.
We associate to each event an energy \eiso\ according to considerations reported in the next section.

\subsection{$E_{\rm \gamma,iso}$ distribution}
\eiso\ is defined as the isotropic equivalent energy emitted in the prompt phase as inferred by an on-axis observer (i.e., an observer whose line-of-sight lies within the jet's core). 
The \eiso\ distribution for the population of short GRBs is built based on the population study presented in \cite{ghirlanda16}. 

Following \citet{ghirlanda16}, we first assign to each event a peak energy $E_{\rm peak}$, randomly extracted from the distribution given in their equation 13. For the values of the function parameters we adopt the mode values for model $a$ reported in their table 1. In this case, the $E_{\rm peak}$ distribution is described by:


\begin{widetext}
\begin{equation}
\phi(E_{\text{peak}}) \propto \begin{cases}
\left(\frac{E_{\text{peak}}}{1.4\,\text{MeV}}\right)^{-0.8} & \text{for}\enspace 0.1\,\text{keV} < E_{\text{peak}} < 1.4\,\text{MeV} \\
\left(\frac{E_{\text{peak}}} {1.4\,\text{MeV}}\right)^{-2.6} & \text{for}\enspace 1.4\,\text{MeV} < E_{\text{peak}} < 100\,\text{MeV}
\end{cases}
\end{equation}
\end{widetext}

For a given value of $E_{\rm peak}$ the associated value of \eiso\ is calculated using the Amati correlation \textbf{for short GRBs} reported in \citealt[(equation 15, with parameters taken from model (a), table 1, mode values)]{ghirlanda16}: 
\begin{equation}
log_{10}(E_{\rm \gamma,iso}/10^{51}\,\rm erg) = 0.036 + 1.1\, log_{10}(E_{\rm peak}/670\,{\rm keV})
\end{equation}

\begin{figure}
    \includegraphics[width=0.98\linewidth]{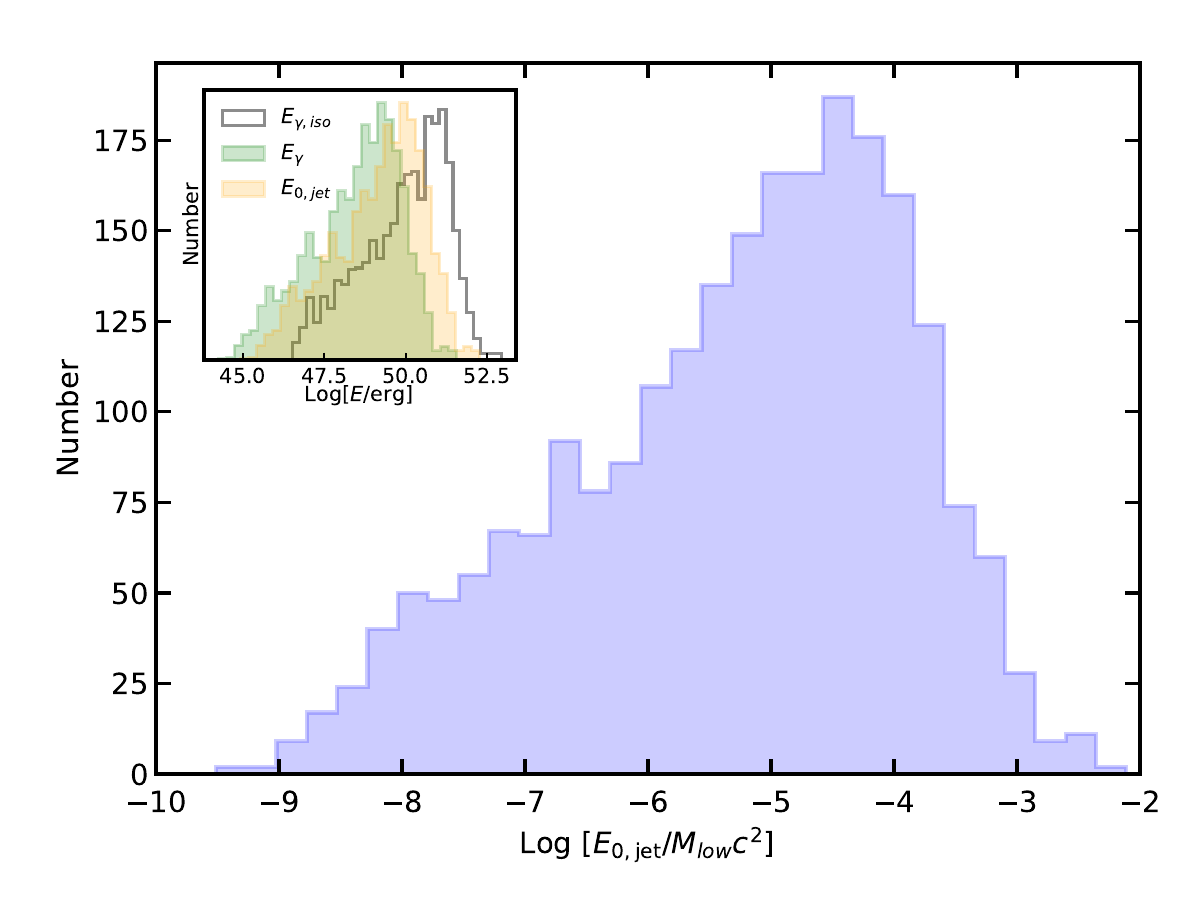}
    \caption{Ratio between the energy of the jet launched following the merger of the BNS and the mass of the lightest NS. The inset shows the distributions of the jet energy $E_{\rm 0,jet}$ (orange filled histogram), the radiated energy $E_\gamma$ (green filled histogram), and the isotropic equivalent radiated energy \eiso\ (black empty histogram).}
    \label{fig:jet_efficiency}
\end{figure}

Once \eiso\ has been assigned, the collimation corrected prompt energy $E_\gamma$, the kinetic energy of the jet after the prompt emission $E_{\rm k,core}$ and the initial jet energy $E_{\rm 0,jet}$ can be computed (see the previous section). Their distributions are shown in the inset of Fig.~\ref{fig:jet_efficiency}, where a value of $\eta_\gamma=0.2$ has been assumed for the radiative efficiency.

As a consistency check, we estimate the efficiency in converting the mass of the disrupted NS into energy of the jet, \textbf{$\eta_{\rm jet}\equiv E_{\rm 0,jet}/M_{\rm low}c^2$}. The distribution of this quantity is shown in Fig.~\ref{fig:jet_efficiency} for which we note the agreement with the typical expected values \citep{colombo_multi-messenger_2025, salafia_efficiency_2021}.

\subsection{VHE lightcurves and spectra}
\label{sec:VHE_emission_lightcurves}
To reproduce the range of spectral behaviours observed in VHE detected GRBs, photon indices are extracted from a Gaussian distribution peaked at $\alpha=-2.2$ and with standard deviation $\sigma=0.1$.

To build the VHE lightcurves, we base our method on the similarity between VHE and soft X-ray afterglow lightcurves.
As shown in several studies \citep{nousek06,kaneko07,davanzo12,margutti13}, the luminosity of the X-ray afterglow (typically estimated in the band 0.3-10\,keV) correlates with \eiso, and this is observed both in long and short GRBs. In most cases, this correlation is studied with reference to the X-ray luminosity measured at 11\,hours. 
Adopting the relation for short GRBs found by \citet{berger14}:
\begin{equation}
L_{\rm X,11h} = 8.5 \times 10^{43} E_{\rm \gamma,iso,51}^{0.83}\,\rm{erg\,s^{-1}}\ ,
\end{equation}
we estimate \lx\ from \eiso. To account for the observed dispersion around the best fit line, we consider a Gaussian scatter of half an order of magnitude.

\begin{figure}
    \centering
    \includegraphics[width=1\linewidth]{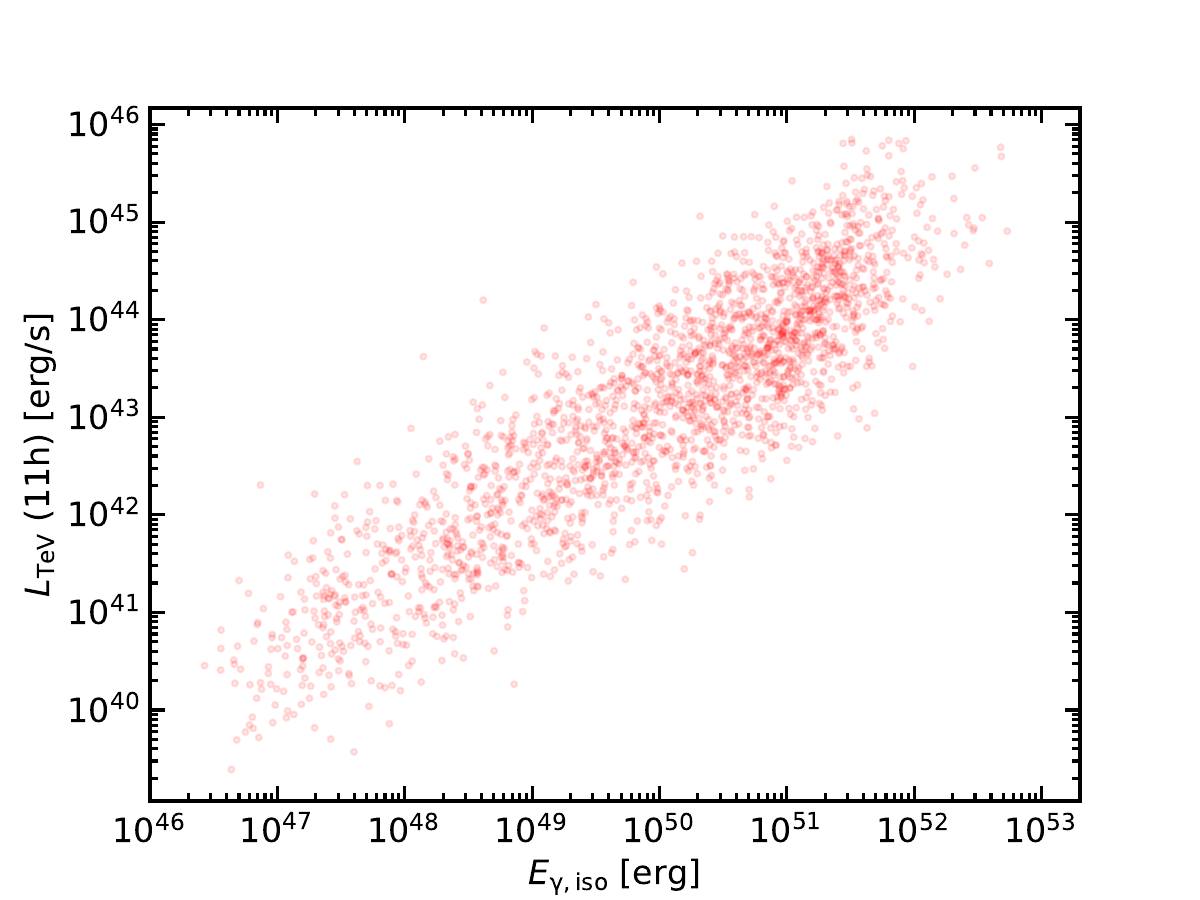}
    \caption{TeV luminosity at 11\,h versus \eiso\ for the sample of short GRBs simulated in this work.}
    \label{fig:ltev-eiso}
\end{figure}

Based on the observation that, for long GRBs, the X-ray and VHE luminosities are similar, the VHE luminosity at 11\,h (\ltev) integrated in the energy range $0.3-1$\,TeV of our sample of simulated short GRB afterglows is computed assuming that the ratio \ltev/\lx\ has a lognormal distribution peaked at 0 and with standard deviation $\sigma=0.3$. 
The resulting correlation between \ltev\ and \eiso\ is shown in Fig.~\ref{fig:ltev-eiso}.
The scatter plot of \ltev\ and \lx\ together with their distributions are shown in Fig.~\ref{fig:luminosities}.

To build the rest of the VHE lightcurve we proceed as follows. We assume that the luminosity increases as a power-law $L_{\rm TeV}\propto t^{\beta_1}$ up to the deceleration time, when it reaches its maximum. The value of $\beta_1$ is randomly extracted from a Gaussian distribution with $\langle\beta_1\rangle=2$ and $\sigma_{\beta1}=0.05$ \citep{LHAASO_221009A}. 
The deceleration radius is computed as:
\begin{equation}            
R_{\rm dec}(\theta) = \left(\frac{17\,E_{\rm k}(\theta)}{16\,\pi\,m_{\rm p}c^2\,n\,\Gamma^2_{0}(\theta)}\right)^{\frac{1}{3}}
\label{eq:deceleration_radius}
\end{equation}
where the density of the external medium $n$ is assumed to be constant and equal to 0.1\,particle\,cm$^{-3}$.

After the peak, during the deceleration, the luminosity decreases as $L_{\rm TeV}\propto t^{\beta_2}$.
The value of $\beta_2$ is assigned after the assumption that VHE and X-ray lightcurves decay at a similar rate, in analogy with what broadly observed in long GRBs. 
We collected a sample of 22 short GRBs and fit their lightcurves with a power-law function (See appendix \ref{appendix_temporal slopes}). The distribution of temporal decay indices is well described by a Gaussian function
with mean value $\langle\beta_2\rangle=-1.45$ and $\sigma_{\beta1}=0.48$. The temporal decays of VHE lightcurves after the peak are randomly extracted from this distribution.
At this point, the VHE lightcurve for an observer oriented along the jet axis is known.

\subsection{Computation of the off-axis emission}

The emission received by an observer located at an angle $\theta_{\rm view}$ is  estimated following the method proposed in \citet{lamb17}. Each differential element of the jet is identified by spherical coordinates ($\theta$ and $\phi$), with the jet axis located at $\theta=0^\circ$. For each jet element, the evolution of the bulk Lorentz factor is assumed to be constant before the deceleration radius (see eq.~\ref{eq:deceleration_radius}) and to decrease in time after the deceleration \citep{BM76}:
\begin{align}            
\qquad  \Gamma(R,\theta)&=
\begin{cases}
\Gamma_0(\theta) & \text{for}\enspace R < R_{\rm dec}(\theta) \\
\left(\frac{17\,E_{\rm k}(\theta)}{16\,\pi\,m_{\rm p}\,n_0\,c^2\,R^3}\right)^{\frac{1}{2}}& \text{for}\enspace R > R_{\rm dec}(\theta)\\[1.2ex]
\end{cases}          
\end{align}

Following \citet{lamb17}, first we estimate the flux received at each observer time by an on-axis observer and then apply corrections to estimate the contribution of each element to the observer located at $\theta_{view}$. Finally, the emission received by the observer is obtained at each time by integrating over all the contributions from different elements of the jet arriving at the same time.

\begin{figure*}
    \centering
    \includegraphics[trim=0 0 30 30,clip,width=0.47\linewidth]{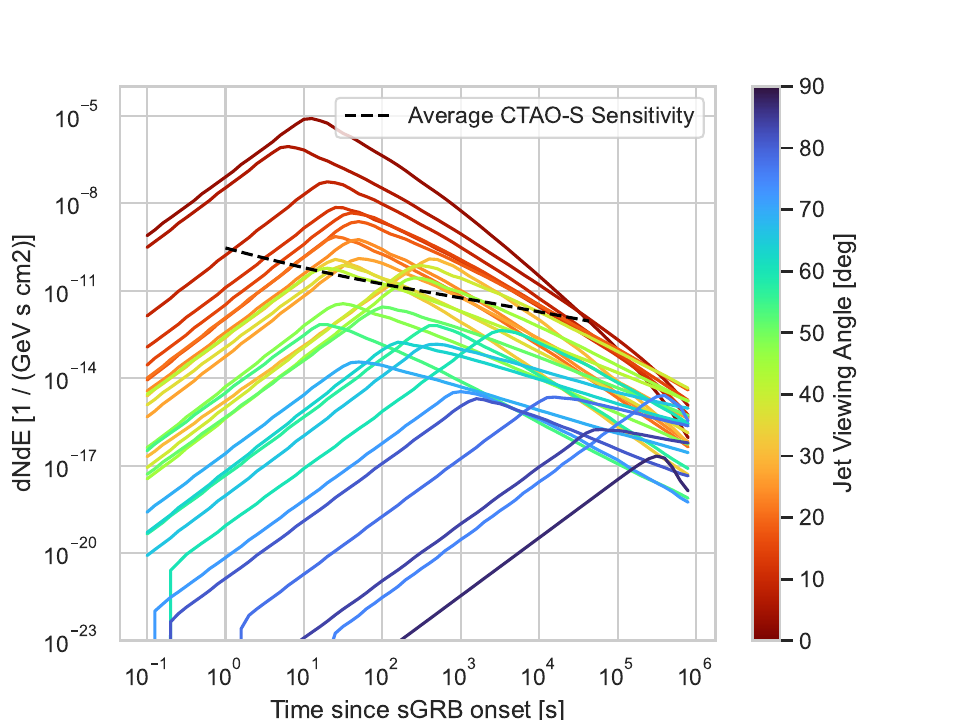}
    \includegraphics[trim=-30 0 10 0,clip,width=0.45\linewidth, height=0.34\linewidth]{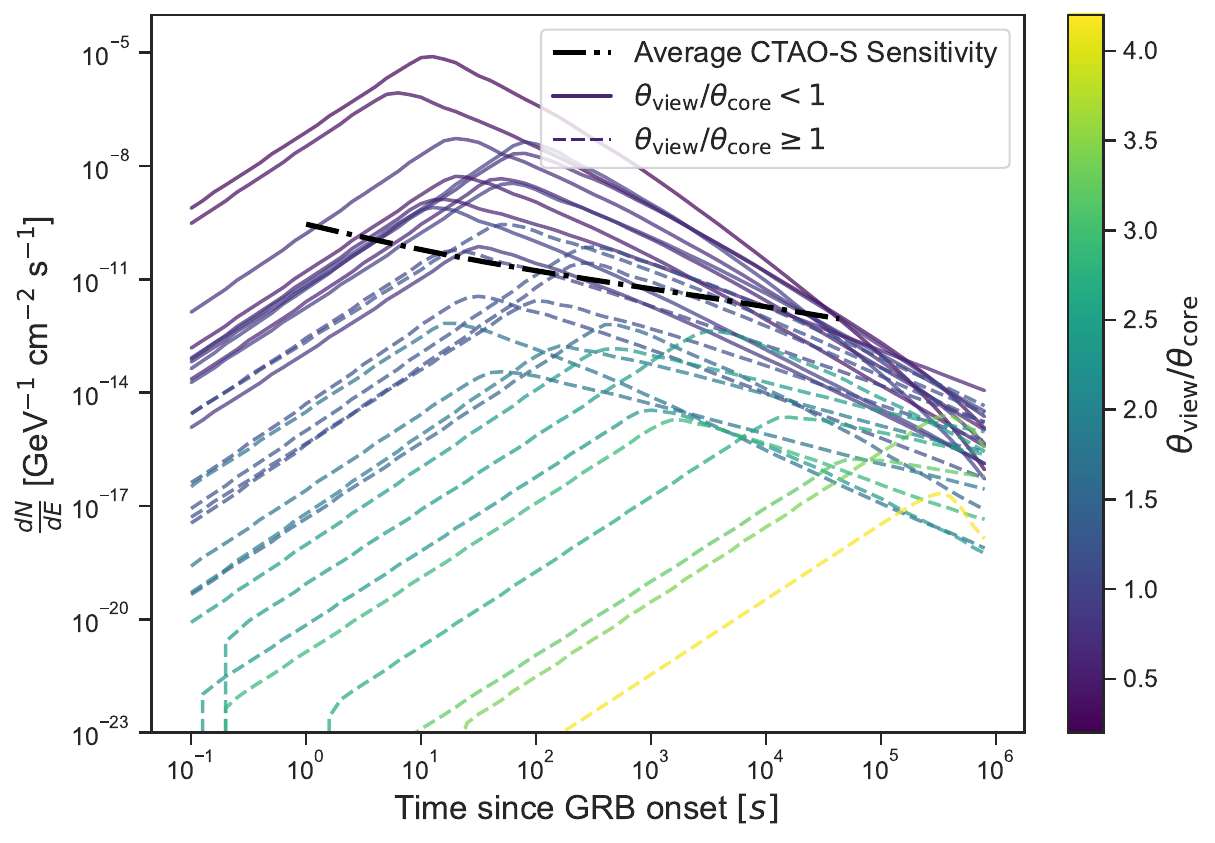}
    \caption{Subset of sGRB lightcurves at 100~GeV generated using the procedure described in Sec~\ref{sec:VHE_emission}. 
    {\em (Left panel)} The color scale corresponds to the off-axis viewing angle between jet and observer, $\theta_{view}$. The dashed, black line indicates the average sGRB integral flux sensitivity of the CTAO-South Alpha configuration (See Sec~\ref{sec:CTAO}), for an exposure time equal to the scale of the x-axis.
    {\em(Right panel)} Subsample of GRB lightcurves color-coded by their respective ratio of $\theta_\mathrm{view} / \theta_\mathrm{core}$. Events for which $\theta_\mathrm{view} / \theta_\mathrm{core} \gt 1$ are shown with dashed lines. The CTAO-South sensitivity is shown for comparison, calculated using the average spectral shape of events in the subsample.}
    \label{fig:lcs}
\end{figure*}

A randomly selected sample of simulated light curves is shown in Fig.~\ref{fig:lcs}. The figure illustrates that the GRB viewing angle has a dual effect on the gamma-ray light curves, shifting them toward both lower spectral fluxes and later times after the sGRB onset. This emphasizes the viewing angle as a key parameter influencing the detectability of a sGRB event. The flux is highest when the observer lies within the opening angle of the jet core, i.e. when the ratio $\theta_\mathrm{view} / \theta_\mathrm{core} \lt 1$, as shown in Figure~\ref{fig:lcs} (right panel). Comparing the light curves with the average CTAO-South sensitivity further highlights the role of the viewing angle in determining the detectability of a given event.

\section{The Cherenkov Telescope Array Observatory in the time-domain astronomy} \label{sec:CTAO}

CTAO is the next-generation ground-based $\gamma$-ray observatory, currently under construction. It will be composed by two arrays of IACTs, one located in the northern hemisphere (CTAO-North, Observatorio Roque de los Muchachos, La Palma, Spain), and one in the southern hemisphere (CTAO-South, Paranal, Chile), which together will provide full sky coverage. The arrays consist of a different combination of telescope designs, which will guarantee a wide energy coverage: the design of CTAO-North includes a total of 4 LSTs (optimized for the 20 GeV - 1 TeV energy range) and 9 Medium Size Telescopes (MSTs, optimized for the 100 GeV - 50 TeV energy range), while CTAO-South will consist of 14 MSTs and 37 Small Size Telescopes (SSTs, optimized for the 1 TeV - 300 TeV energy range). This configuration is named "Alpha-configuration".

CTAO will play a crucial role in $\gamma$-ray astronomy due to its unprecedented sensitivity, up to an order of magnitude better than current instruments, its rapid slewing capabilities (e.g. the LSTs can be re-pointed in about 20 seconds), its large FoV ranging from 4.3$^{\circ}$ to 10$^{\circ}$ depending on the telescope, and angular resolution (down to few arcminutes) \citealp{2019scta.book.....C}.
In addition, CTAO will have a very high sensitivity to short-timescale phenomena: at energies above few tens of GeV, it offers 10$^4$-10$^5$ better sensitivity than the LAT instrument onboard the Fermi satellite for the detection of short-duration transient phenomena (\citealp{2013APh....43..348F}; see also \href{https://www.ctao.org/for-scientists/performance/}{https://www.ctao.org/for-scientists/performance/}).

\subsection{The ingestion of GW alerts with CTAO}

CTAO features the ACADA system (Array Control and Data Acquisition,  \citealp{oya2024first}), which coordinates telescope operations, data acquisition, and real-time data processing, executing pre-scheduled observations and those triggered by science alerts. Science alerts can be triggered externally (e.g. through GCN or VoEvent), or internally thanks to the Science Alert Generation (SAG) pipeline \citep{SAG2022}, which processes incoming Cherenkov data, reconstructs events, and generates science alerts within ~20–30 seconds at nominal CTAO sensitivity.

External and internal alerts are ingested by the Transient Handler (TH), a sub-system of ACADA \citep{TH2022}. The TH validates the alerts, applies configurable science filters, and, when warranted, triggers, updates, or retracts follow-up observations.

In the context of responding to GW alerts, the TH will play a central role in validating and selecting the numerous alerts expected during O5, and in implementing the optimal observation strategy. This includes selecting the ensemble of observation coordinates required to cover to cover a significant localization region, scheduling observing times, and configuring sub-groups of telescopes (i.e. sub-arrays) to maximize scientific output within the constraints of the GW alerts and based on current knowledge of the expected gamma-ray counterparts. More details will be provided in section \ref{sec:obs-strategy}. 

Further, the SAG enables the evaluation of acquired data for early identification of transient phenomena and the generation of internal candidate scientific alerts, submitted to the TH. To this end, and because sky area is tiled with multiple pointings, data stacking across multiple observations is essential. In cases involving large sky regions, the system developed by ACADA ensures accurate stacking over consistent coordinates throughout the entire observation campaign, even in the presence of observing gaps.

\subsection{CTAO telescopes configuration and Instrument Response Function}

The performance characterization of the CTAO is based on detailed Monte Carlo simulations, leading to the generation of Instrument Response Functions (IRFs). These IRFs quantify the sensitivity, effective area, angular and energy resolution, and background rates of the array as functions of energy, observation direction, and offset angle. 

The current baseline layout adopted for CTAO is known as the Alpha Configuration, and was described in the previous subsection. The IRFs associated with this configuration were derived from the Prod5 simulations\footnote{\url{https://doi.org/10.5281/zenodo.5499840}}. These use CORSIKA (v7.7) and sim-telarray toolchains, incorporating updated models of atmospheric properties, telescope optics, and electronics\footnote{\url{https://doi.org/10.5281/zenodo.5499840}}, and represents a substantial advancement over previous simulation sets. 
Because performance is heavily dependent on pointing, the Prod5 IRFs contain performance estimates for both the Northern and Southern arrays at three different zenith angles: $20^\circ$, $40^\circ$, and $60^\circ$. The lower energy thresholds adopted for each zenith are $E_\text{min}$ used in this work are $20$~GeV, $32$~GeV, and $130$~GeV, respectively. The maximum energy $E_\text{max}$ adopted is $10$~TeV for all IRFs.

\section{Simulation of CTAO observation and analysis}\label{sec:CTAsim}

The VHE emission derived in the previous sections is used as input emission for which the CTAO detectability prospects are evaluated. We include the attenuation of the intrinsic GRB spectrum by the EBL, as modeled in \citet{franceschini_ebl}, to obtain the observed flux for each GRB. For the simulation of the CTAO response, the CTAO Alpha Configuration is considered, as described in Section \ref{sec:CTAO}.

In these simulations, the exposure time required for CTAO to detect each event is estimated as a function of the delay time from the onset of the GRB emission. This approach allows for the evaluation of each GRB's detectability within the parameter space defined by exposure time and delay. The minimum fluence detectable by CTAO is computed using the IRFs (see Section~\ref{sec:CTAO}), and the exposure time is calculated as the time required for a 5$\sigma$ detection by CTAO. All significance values and detection criteria in the following sections refer to pretrial significance, unless stated otherwise.

\subsection{Simulating CTAO Response to BNS Merger Events}
The following step in our pipeline is to simulate the response of CTAO to each of the GRB events produced as described in Section \ref{sec:VHE_emission}. The ultimate goal of these simulations is to find, for each event, the functional relationship between latency from the onset of the merger and the total observation time needed to achieve a $5\sigma$ detection level with CTAO. Given this relationship, one can quickly perform a lookup to check how much observation time is needed to detect an event given the current latency. This analysis yields several insights, including the average detectability rate across all simulated events and the influence of GRB and jet properties on detectability. In addition, these results can be provided as input to help optimize real observing strategies (see Section \ref{sec:obs-strategy}).

Observations of the correct source position typically start after a certain latency $t_L$ from the BNS merger time $t_0$: $t_\text{start} = t_0 + t_L$. $t_L$ has contributions from many stacking factors, including the time for the LVK collaboration to process and emit a GW alert, as well as the TH handling and the telescope slewing time. In addition, a number of different pointings may be necessary before the true source location is pinpointed, each of which contributes to $t_L$.

The strategy for calculating this relationship relies on comparing the flux needed to detect the EM counterpart at a $5\sigma$ level with CTAO, $F^{\text{CTAO}}_{5\sigma}$, with the average flux of the source $F_\text{avg}$. The time needed for detection $t_\text{det}$ is defined as the exposure time ($t_\text{exp}$) that satisfies the following equality:

\begin{equation}            
    F_{\text{avg}}(t_\text{exp}; t_L) = 
    F^{\text{CTAO}}_{5\sigma}(t_\text{exp})
\label{eq:gw-toy-eq}
\end{equation}

The average flux is given by:

\begin{equation}            
F_{\text{avg}}(t_\text{exp}; t_L) = \frac{S(t_\text{exp}; t_L)}{t_\text{exp}} \; ,
\end{equation}
where $S$ is the intrinsic source fluence:
\begin{equation}            
S(t_\text{exp}; t_L) = \int_{t_L}^{ t_L + t_\text{exp}} \int_{E_\text{min}}^{E_\text{max}} \phi(E, t)\ E \ dE \ dt \; ,
\end{equation}
where $\phi(E, t)$ represents the GRB spectra as described in Section \ref{sec:VHE_emission}, and $E_\text{min}$ and $E_\text{max}$ correspond to the energy limits of the CTAO IRFs (Section~\ref{sec:CTAO}). In the above equations, we assume that the GRB onset time is simultaneous with BNS merger time, such that we can explicitly set $t_0 = 0$.

These simulations are performed with a custom-tailored Python package called \texttt{sensipy} \citep{green_sensipy_2026}, which takes CTAO IRFs, an EBL model~\citep{franceschini_ebl}, and GRB simulations as input and calculates detection times given a variety of initial conditions and user inputs, essentially by solving Equation \ref{eq:gw-toy-eq}. For the management of spectral models, the package depends on primitives from \texttt{gammapy}, the analysis package used by CTAO~\citep{gammapy_2023}.

\begin{figure}[tb]
    \centering
    \includegraphics[width=0.45\textwidth]{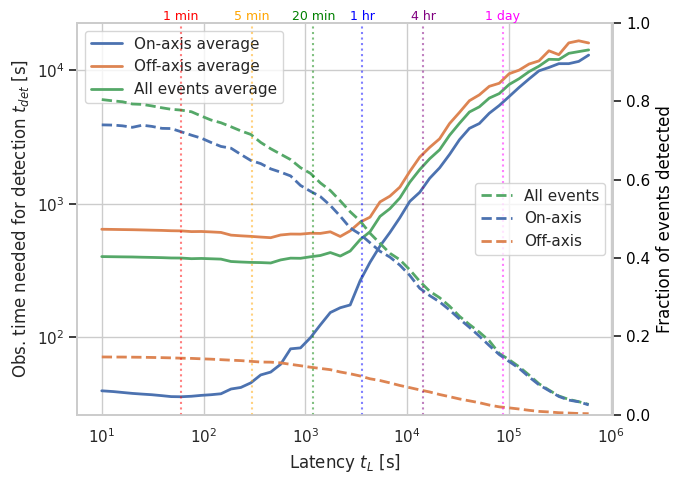}
    \caption{All curves in this plot are calculated using only events which are detectable by CTAO within \SI{7}{\day} of GRB onset. For these detectable events, the solid curves show the average observation time needed to reach the $5\sigma$ significance threshold given a latency $t_L$ between GW onset and start of observations. For observing latencies of $t_L \lesssim$ \SI{1}{\hour}, nearly the entire population of detectable GRBs can be detected with observations of 1-\SI{10}{\min} exposures. After this threshold, the viewing angle of the GRB is not an important contributing factor to the observation time needed for detection. The dashed curves show the fraction of events which are detectable given $t_L$. Once again, after the $1$~h threshold, the fraction of detections descends from a constant 20\% to nearly 0\% after \SI{24}{\hour}. All curves show the relative importance of the first few hours post-trigger.}
    \label{fig:gw-toy-flux}
\end{figure}

\subsection{CTAO Response to Simulated GRBs}
\label{sec:ctao-response-to-sgrbs}

We begin with the simulated CTAO response to the 2307 GW-GRB events to estimate the impact of observing time on $5\sigma$ detectability. Each GRB is simulated with six different CTAO IRFs: both sites, and at three different zenith angles (\SI{20}{\degree}, \SI{40}{\degree}, and \SI{60}{\degree}). In addition, each event is observed with each IRF at 50 different $t_L$, distributed on a logarithmic scale from \SI{10}{\sec} to \SI{7}{\day}, leading to a total catalog size of nearly \SI{7e5} simulations.
Figure \ref{fig:gw-toy-flux} shows the average relationship between $t_L$ and $t_{det}$ only for events that are detectable by CTAO within \SI{7}{\day} of the GRB onset, about $\sim16$\% of the total sample. These curves show how the average exposure time needed for a $5\sigma$ detection increases with the increasing time delay between the GW-GRB onset and the start of observations at the true source location. In the same figure, we also show the fraction of events detectable with the average exposure time at the given latency. As expected, the fraction of detectable events decreases with increasing latency. In addition, with a latency lower than $\sim 1$~hr, the exposure time necessary for detection remains more or less constant. Similarly, the fraction of detected events also decreases after a similar latency time. Lastly, in comparing the results for on- and off-axis sub-samples at low latencies, on-axis GRBs can be detected within a few minutes on average, whereas off-axis GRBs require $\sim10$~min for detection. Above this same threshold of \SI{\sim1}{\hour}, the viewing angle of the event becomes much less important, and the detection time scales exponentially with the latency for all events. These results emphasize the fact that, as observed for long GRB, the first minutes to hours are critical for detecting GW-GRB coincidences~\citep{2019Natur.575..455M,MAGIC_201216C,2019Natur.575..464A,2021Sci...372.1081H}.

\begin{figure*}[tb]
    \centering

    \begin{subfigure}{0.45\textwidth}
        \includegraphics[width=\linewidth]{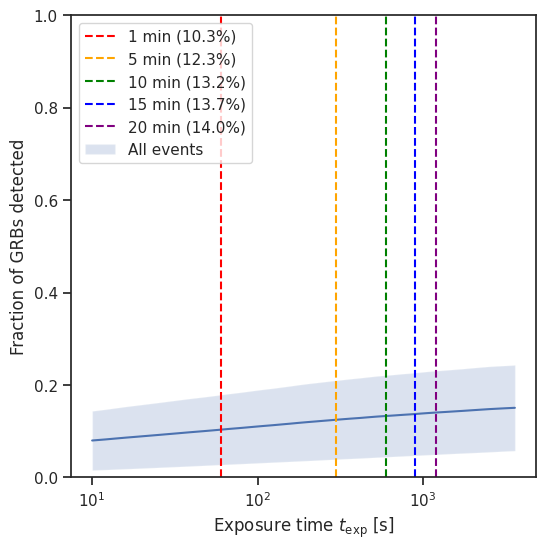}
        \caption{Fraction of all events detected with increasing exposure time}
        \label{fig:cta-south-exposure-vs-prob}
    \end{subfigure}
    \hfill
    \begin{subfigure}{0.45\textwidth}
        \includegraphics[width=\linewidth]{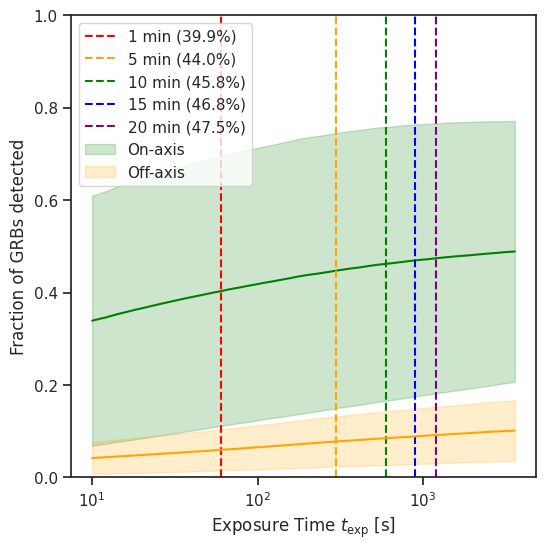}
        \caption{Fraction of on- and off-axis events detected with increasing exposure time}
        \label{fig:cta-south-exposure-vs-prob-axis}
    \end{subfigure}

    \vskip\baselineskip

    \begin{subfigure}{0.45\textwidth}
        \includegraphics[width=\linewidth]{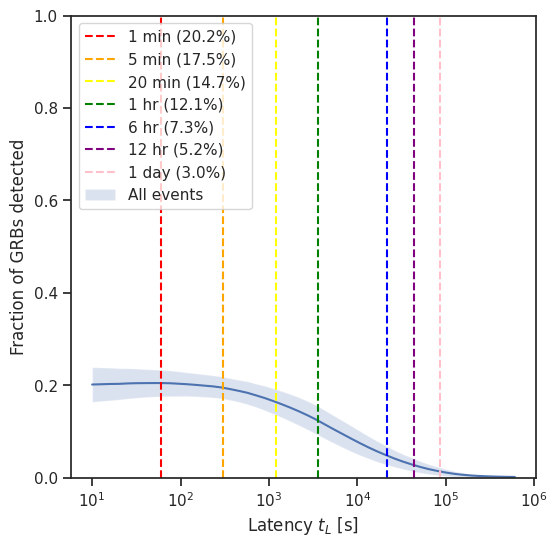}
        \caption{Fraction of all events detected with increasing latency after GRB onset}
        \label{fig:cta-south-latency-vs-prob}
    \end{subfigure}
    \hfill
    \begin{subfigure}{0.45\textwidth}
        \includegraphics[width=\linewidth]{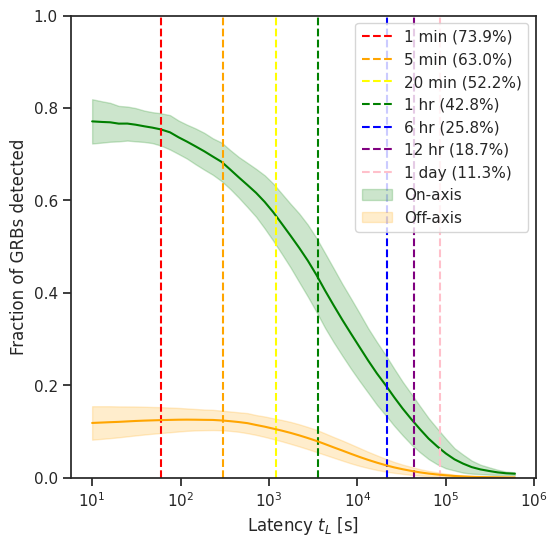}
        \caption{Fraction of on- and off-axis events detected with increasing latency after GRB onset}
        \label{fig:cta-south-latency-vs-prob-axis}
    \end{subfigure}

    \caption{These curves explore the relationship between the fraction of events detected and two follow-up observation parameters: exposure time on source position $t_\text{exp}$ and latency time after coalescence $t_L$. The upper left and right panels show how increasing exposure leads to a cumulative increase in detections, for all events and on- and off-axis events, respectively. In all cases, increasing exposures beyond $t_{exp} \gtrsim 10$~min leads to only a small increase in the fraction of detected events from the sample. The lower panels show how increasing $t_L$ for all, on-, and off-axis events lead to a vanishing fraction of detectable GRBs. On average, the fraction of detected events decreases fourfold after latencies of $\sim 12$~hrs. In this sample, each event is simulated 300 times, at different values of zenith, CTAO site, and with $t_L$ ranging from \SI{10}{\sec} to \SI{7}{\day}. The large uncertainty bands around some curves are due to small samples sizes of detectable events.}
    \label{fig:cta-south-projection-plots}
\end{figure*}

Another way to interpret these simulations is to consider a specific value of $t_L$ or $t_{exp}$, and estimate the probability of detecting a certain event. As a concrete example, a GW-followup scheduler must make an informed decision when no signal is detected yet: keep observing the same position or move on to the next pointing? Figure~\ref{fig:cta-south-exposure-vs-prob} and Figure~\ref{fig:cta-south-exposure-vs-prob-axis} illustrate that, in most cases, significantly increasing the exposure time does not increase the chances of detection. Extending an observation from $1$~min to $20$~min only increases the overall probability of detection from $10.3\%$ to $14\%$. In addition, Figure~\ref{fig:cta-south-latency-vs-prob} and Figure~\ref{fig:cta-south-latency-vs-prob-axis} indicate that latencies longer than a few hours significantly decrease the probability of detection. In most cases, continuing observations beyond $\sim6$~hr presents a significant challenge for joint detections. The large difference in detectability between on-axis and off-axis events is driven by the connection between spectral flux and $\theta_{view}$, as noted in Figure~\ref{fig:lcs}. In addition, we observe that observations with a latency of $24$~h may still be fruitful. Overall, the plots in Figure~\ref{fig:cta-south-projection-plots} suggest that, under the assumption that the simulated population of BNS mergers well-represents the true distribution, on-axis events offer a significant opportunity for a joint detection with CTAO. 

\begin{figure*}[tb]
    \centering

    \begin{subfigure}{0.45\textwidth}
        \includegraphics[width=\linewidth]{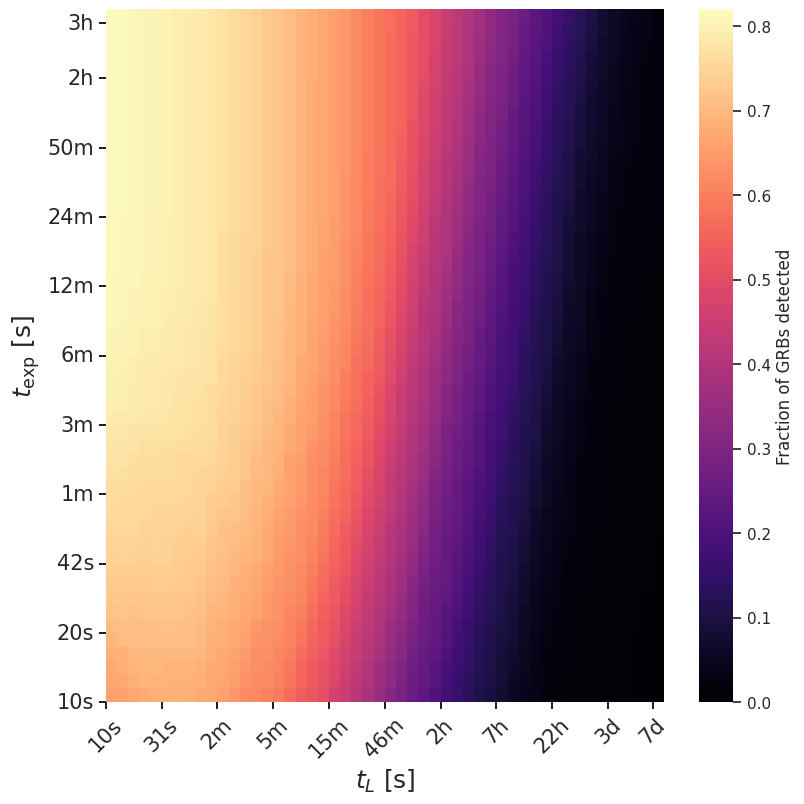}
        \caption{CTAO-South on-axis}
        \label{fig:heatmap-south-onaxis}
    \end{subfigure}
    \hfill
    \begin{subfigure}{0.45\textwidth}
        \includegraphics[width=\linewidth]{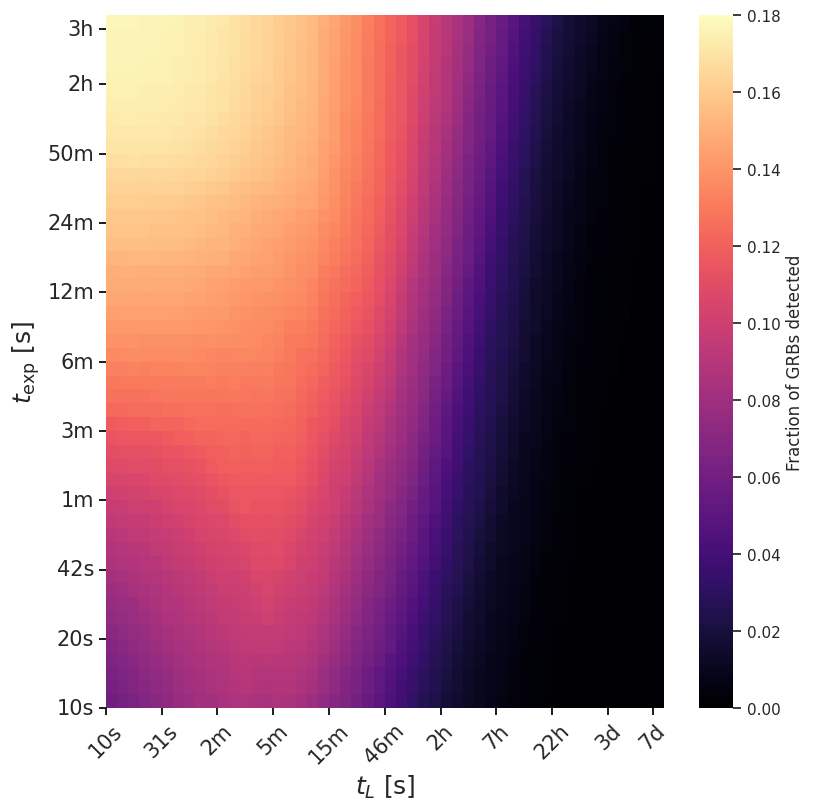}
        \caption{CTAO-South off-axis}
        \label{fig:heatmap-south-offaxis}
    \end{subfigure}

    \vskip\baselineskip

    \begin{subfigure}{0.45\textwidth}
        \includegraphics[width=\linewidth]{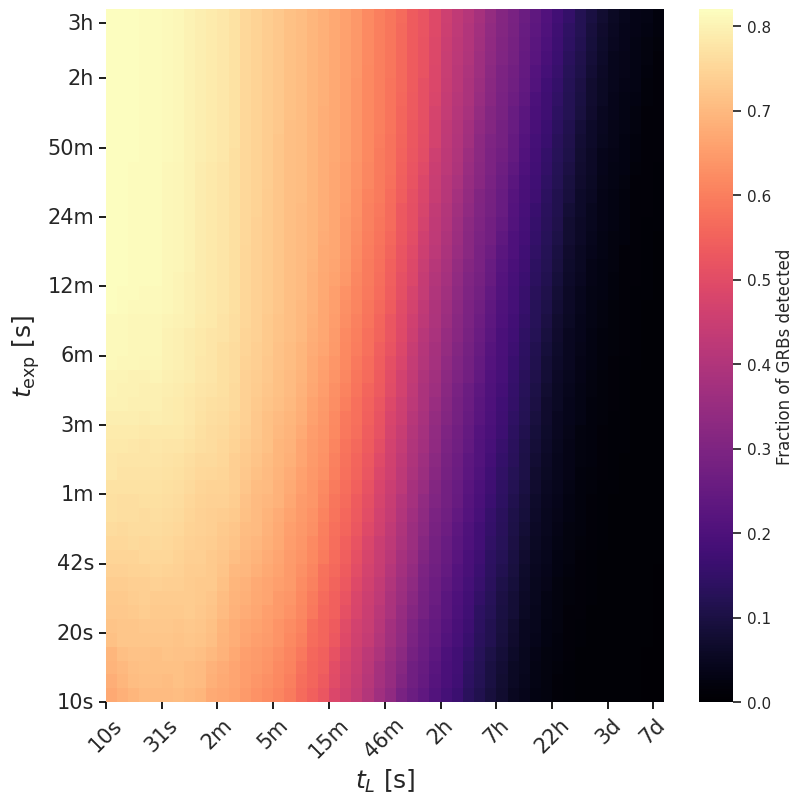}
        \caption{CTAO-North on-axis}
        \label{fig:heatmap-north-onaxis}
    \end{subfigure}
    \hfill
    \begin{subfigure}{0.45\textwidth}
        \includegraphics[width=\linewidth]{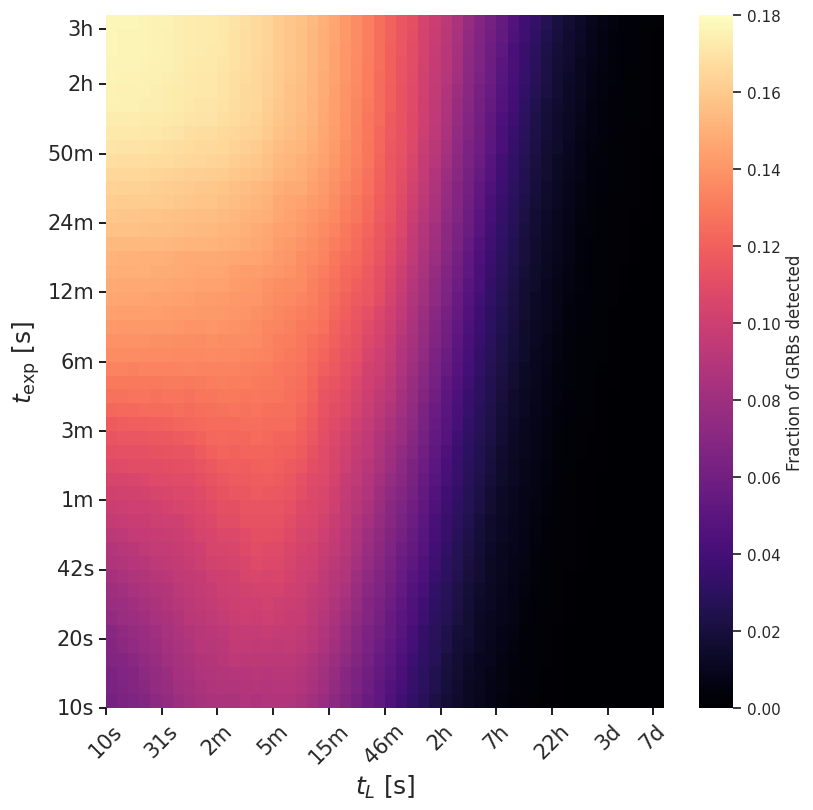}
        \caption{CTAO-North off-axis}
        \label{fig:heatmap-north-offaxis}
    \end{subfigure}

    \caption{Detectability heatmaps for short gamma-ray bursts using CTAO South (upper row) and CTAO North (bottom row) across the complete simulated dataset, shown as functions of latency $t_{L}$ and exposure time $t_{\text{exp}}$. The left column displays results restricted to on-axis sources with viewing angles within the opening angle of the jet (namely having $\theta_{\text{eff}} < 0$), whereas the right column encompasses all sources within $\theta_{\text{eff}} > 0$. The color scale represents the proportion of simulated events successfully detected, with brighter tones indicating enhanced detection efficiency. Notably, we note that the largest detectability change due to the observation latency appears between ~1-4~h, which ultimately depends on the observation exposure.}
    \label{fig:heatmaps}
\end{figure*}

\begin{figure*}[htb]
    \centering
    \includegraphics[width=0.8\textwidth]{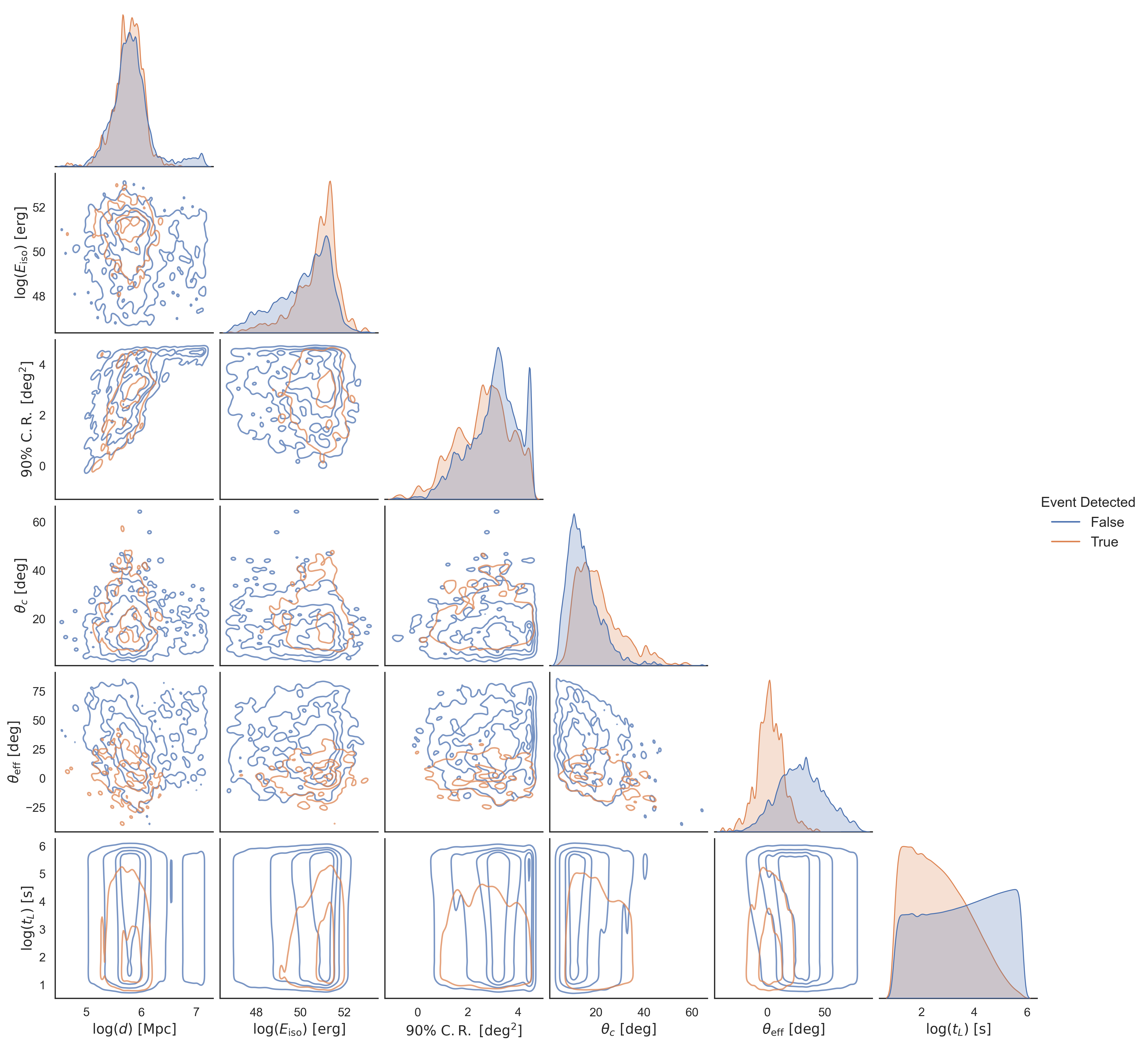}
    \caption{Relationship between pairs of intrinsic GRB and observational parameters. Blue points represent detectable events from our sample while orange points are non-detectable events. $\theta_{core}$ is the opening angle of the cone of the GRB jet, and $\theta_{\text{eff}}$ is the angular distance between the observer and the edge of the cone. The strongest distinction between the distributions of detectable and non-detectable events seen in $\theta_{\text{eff}}$, suggest that the single most important factor contributing to detectability is whether or not the observer lies within or near the edge of the jet. In addition, we see once more that with latencies $t_L$ under a few hours, we have a higher number of simulated events which are detectable.}
    \label{fig:corner-plot}
\end{figure*}

In order to obtain a more holistic view of the $t_L$-$t_{exp}$ parameter space, in Figure~\ref{fig:heatmaps} we plot both variables together against the fraction of detected events in the sample. Here, we define a new parameter that represents the effective viewing angle of the source $\theta_{\text{eff}} = \theta_{\text{view}} - \theta_{core}$, where $\theta_{\text{view}}$ is the viewing angle of the observer. When $\theta_{\text{eff}} < 0$, the observer is on axis and therefore subject to viewing the beaming effect of the GRB jet, whereas off-axis events are defined by $\theta_{\text{eff}} > 0$, dominating our population of GW-GRBs, being 87\% of the total population. 

Two distinct regions of detectability are visible, separated by a steep curve in latency space. For both on- and off-axis GRBs, the detectability exhibits a pronounced decline for an observational latency of approximately~1hr, with the precise latency threshold at which this decrease occurs being dependent of the exposure. These heatmaps also show that for $t_L = 10$~min, approximately 67\% of on-axis sources can be detected with an exposure time of $\le1$~min. For a longer latency of $t_L = 1$~hr, the fraction of detected events is 47\% with the same exposure time of $\le1$~min. For off-axis events, these rates go down to 11\% and 7\% for $t_L = 10$~min and $t_L = 1$~hr, respectively. It can also be seen that detections are only slightly more likely at CTAO-South compared to CTAO-North, which we interpret as resulting from the larger number of MSTs. We obtained an overall detection rate of 18.3\% and 17.6\%, respectively. Lastly, contour lines on these heatmaps indicate a potential strategy for IACT followup campaigns: pick a constant value of probability from such heatmaps and use these to calculate the exposure time at each pointing. It is also important to note that these numbers do not yet take into account day and night cycles, the moon, and other observational limits which would affect the true detectability of a given source. These factors will be considered in Section \ref{sec:obs-strategy}. 

\subsection{Linking CTAO Response to GW-GRB parameters}
\label{sec:linking CTAO to GW-GRB}
Given the large sample of events simulated under different observing conditions and values of $t_L$, we can also examine the relationship between the inherent parameters of the simulated GRBs with the CTAO response. The corner plot of Figure \ref{fig:corner-plot} explores the relationship between various observational parameters and detectability. The parameters considered include source distance ${d}$, isotropic energy $E_\text{iso}$, the angular size of the 90\% C.R. of the GW skymap, the opening angle of the GRB jet $\theta_{core}$, the effective viewing angle $\theta_{\text{eff}}$ (see Section~\ref{sec:ctao-response-to-sgrbs}), and the latency $t_L$.

The corner plot indicates that both $\theta_{\text{eff}}$ and $t_L$ have the most distinct populations between detected and undetected events of the sample.While $t_L$ is always known during followup campaigns, GW observatories do not currently provide any estimate on $\theta_{\text{view}}$~\citep{2020LRR....23....3A}. Should they gain such an ability to provide constraints on the orientation of mergers with respect to the observer, this could provide one of the most stringent criteria for whether to carry out a GW followup.

\section{Observing strategies} \label{sec:obs-strategy}
Compared to other transient sources, gravitational wave follow-up observations present an extra level of complexity given the average extension of the source localization uncertainty region. During O4, tens of gravitational wave candidates have been localized down to several (tens of) degrees for their 50\% (90\%) C.R. (e.g. S250119cv \citep{2025GCN.38991....1L} or S241102br \citep{2025GCN38043}, see \href{https://gracedb.ligo.org/superevents/S240615dg}{GraceDB} for the full list). These uncertainty regions can be further reduced as more GW interferometers join the global network and contribute to the GW sky localization.
The LST and MST telescopes of CTAO are designed to cover a large fraction of the VHE energy range where we expect most of the GRB above-GeV emission. The FoV of these telescopes, with FoV$_{radius} \sim 2.5$~deg, means that GW events with a 90\% C.R. smaller than $\sim$ 20 deg$^2$ may be subsequently observed as \textit{point-like} sources, following a \textit{wobble} strategy. Anything larger than this will require a dedicated follow-up strategy.

\subsection{Tiling observations of poorly localized GW}
A substantial number of gravitational-wave detections will exhibit poorly constrained localization, highlighting the importance of observation strategies. Following up such poorly-localized events presents a particular challenge for pointing instruments like CTAO, which require robust localization strategies. In this work, the observation campaign for each GW is handled by \texttt{tilepy}~\citep{seglar2024cross}. \texttt{tilepy} is a software package designed to plan and optimize follow-up observations of transient events with extended sky localizations. It ingests probabilistic sky-localization maps (e.g. HEALPix maps from the LIGO–Virgo–KAGRA network) and generates candidate telescope pointings based on the instrument field of view, site visibility, and observational constraints. The tool optimizes the selection and scheduling of these pointings to maximize the covered localization probability while accounting for latency, exposure time, night duration, and observing conditions such as zenith angle.
In this study, we focus on the temporal observation campaign optimization by studying how this choice impacts the GRB detectability results.  For this purpose, we have identified three different scenarios regarding exposures: fixed exposures, average-GRB emission exposures, and adaptive exposures. 

The first case studied focuses on fixed observation exposures.  We have selected three different exposures, which account to a total of three different cases. These are: 1-minute exposure time (to study the fastest case, for which from Sec. \ref{sec:CTAsim} the source is detected in $\sim$ 73\% of the on-axis cases and $\sim$12\% of the off-axis cases), a 5-minute exposure time for which the source is detected in $\sim$ 63\% of the on-axis cases ($\sim$12\% of the off-axis) and 20-minute exposure time, for which the source is still detected in  $\sim$ 52\% of the on-axis cases ($\sim$10\% of the off-axis). Note that the latest choice coincides with the standard pointing duration in current-generation IACTs, and is therefore adopted in most VHE follow-up campaigns.

The second case concerns exposures modulated according to the temporal evolution of an average GRB emission, derived from the full set of light curves in our sample, which generates a sort of representative source. For every new pointing, we calculate the time necessary to achieve a \(5\sigma\) detection of this representative source, incorporating the pointing latency \(t_L\) (following Equation~\ref{eq:gw-toy-eq}). Crucially, we now treat the exposure time per tiling, typically a static configuration parameter, as an adjustable degree of freedom, necessitating specific adaptations within \texttt{tilepy}.

Lastly, in our third study scenario, we incorporate the known spectral evolution of each GRB when scheduling its pointings. While such a strategy cannot be used during real follow-ups, it can provide a baseline for the efficiency of a campaign. Within this framework, the exact exposure time for each pointing is determined as the period necessary to achieve a \(5\sigma\) signal, accounting for each event's specific light curve and the current latency \(t_L\), precisely as described in Section \ref{sec:VHE_emission}. This stringent requirement ensures, by definition, that the source is always detected, provided that the tiling successfully encompasses its position within the observing constraints. We note that for some events, however, this strategy can lead to large delays before source position coverage, especially compared to the short, fixed-window strategies.

We ran \texttt{tilepy} on the full set of GW-sGRB simulations for the three mentioned scenarios, which account for a total of five configurations. In all cases, a conservative FoV radius of 2.5$^{\circ}$ is selected for the scheduling, corresponding to the radius at which the camera acceptance has decreased by a factor $\sim$2.
This value is intended to be representative of an overall combined FoV of CTAO. The simulation is configured to have a maximum duration of 3 full days. The selected observation windows are realistic. We consider the site visibility conditions, which include Sun position, Moon position (for which we allow for observations under certain moonlight conditions defined by altitude, phase and separation to the observation coordinates) and the evolution of the localization region throughout the night.
Firstly, we note that, for all cases, the observability conditions at the observatory represent a key consideration that modulates the CTAO's duty cycle and these results. These considerations are taken into account to assess the scheduling of the source for the input merger time of the BNS and the GRB evolution.
The detectability of a GRB depends on the zenith angle of the observation, which changes the energy threshold and effective area. The precision of the current study is set by the available IRFs, which are derived for fixed zenith angle values of  20$^{\circ}$, 40$^{\circ}$ and 60 $^{\circ}$ (see Sec~\ref{sec:CTAO}). These are used in our study to obtain the 5$\sigma$ signal condition evaluated for each possible tiling, after a pre-selection based on the probability coverage. For a pointing to be scheduled, its coordinates must be observable by the telescope at both the beginning and end of the proposed window; otherwise the pointing is skipped. This process is repeated until the full observing campaign is assembled. The future CTAO will produce tailored Monte Carlo simulations for the analysis of the acquired data, which will improve the detection prospects of our tiling strategy. 

This approach folds the duty cycle of IACTs in our simulation chain, which reaches  $\sim$15\% per site over a year, including day time. 
 
\begin{figure*}[tb] 
    \centering
    \begin{subfigure}{0.545\textwidth}
        \centering
        \includegraphics[width=0.9\linewidth]{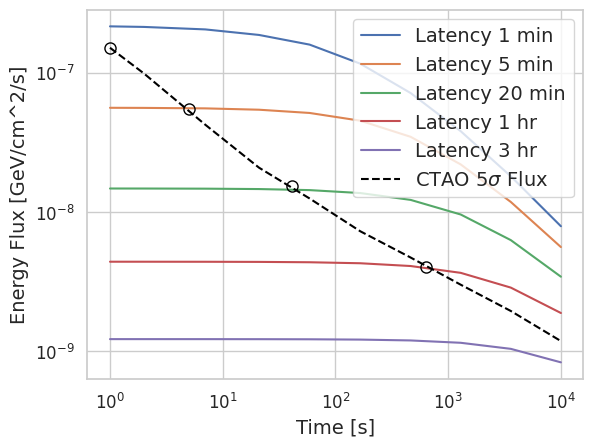}
        \caption{Example GRB lightcurves for one event at different latencies (colored lines). The intersection of these curves with the $5\sigma$ CTAO Flux sensitivity (dashed line) indicates the observation time needed to achieve a detection for each latency. For the lightcurves, the horizontal axis represents time since GRB onset, while for the sensitivity curve it represents observation time $t_\text{exp}.$}
    \end{subfigure}
    \hfill
    \begin{subfigure}{0.444\textwidth}
        \centering
        \includegraphics[width=0.8\linewidth]{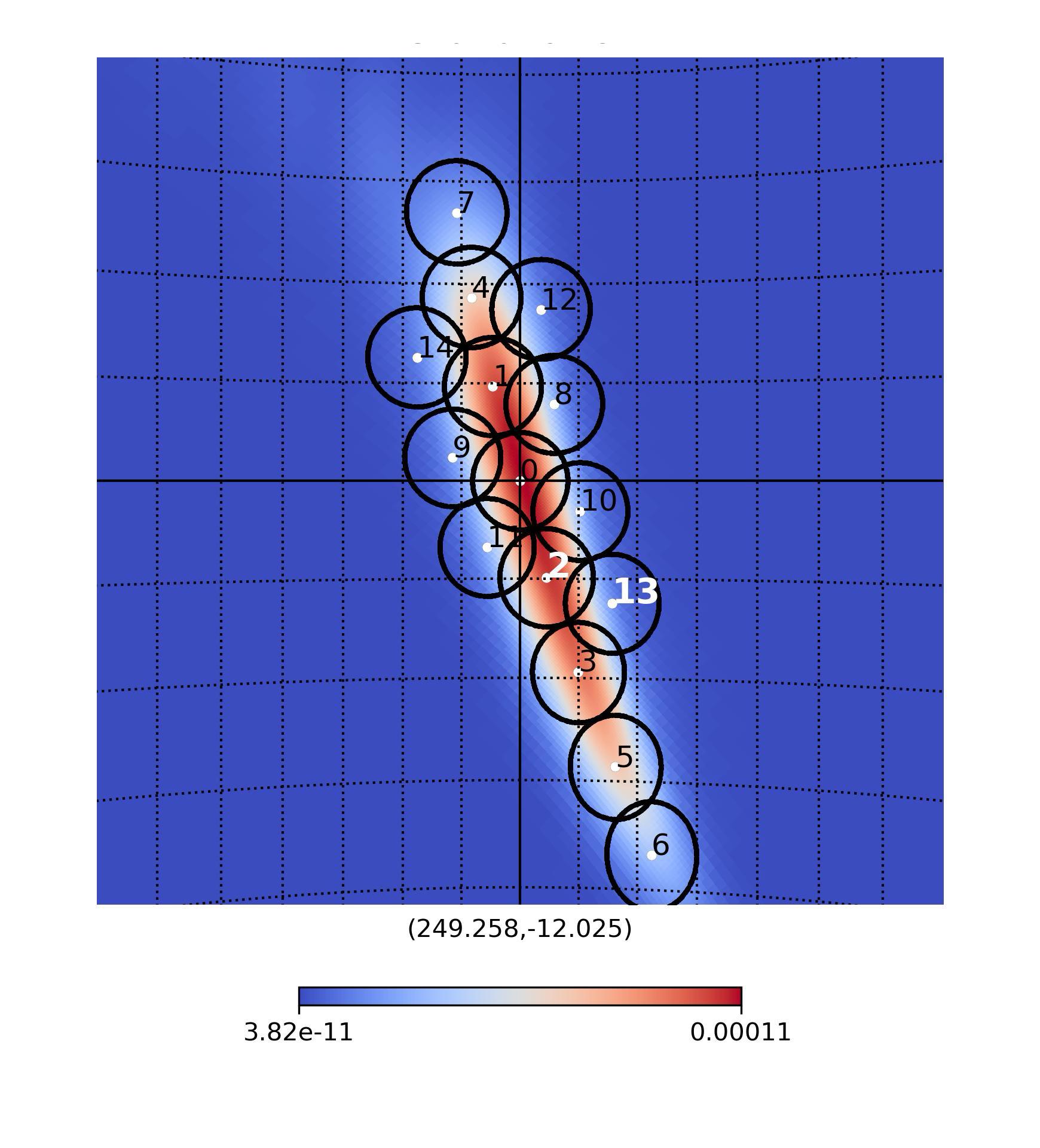}
        \caption{Observation scheduling for the GW skymap associated to the example GW-GRB event. The color scale represents the probability region provided by the interferometer pipeline, and the numbers indicate the sequential ordering of CTAO observations. The source is detected both in pointings 2 and 13, indicated by white numbers.}
    \end{subfigure}
        \caption{Example lightcurves and tiling strategy for a representative event from the sample. This sGRB is seen on-axis, has an $E_{iso} = 1.21\times10^{51}$~erg, and it is located at a distance $d = 991$~Mpc. The EBL is taken into consideration with the models from \cite{franceschini_ebl}. The resulting skymap has a 90\% credible area of $345\deg^2$.}
    \label{fig:tiling-LC}
\end{figure*}

An example of the integration between tiling, CTAO sensitivity and the GRB lightcurves can be found in Figure \ref{fig:tiling-LC}.

\begin{figure}[tb]
    \includegraphics[width=0.98\linewidth]{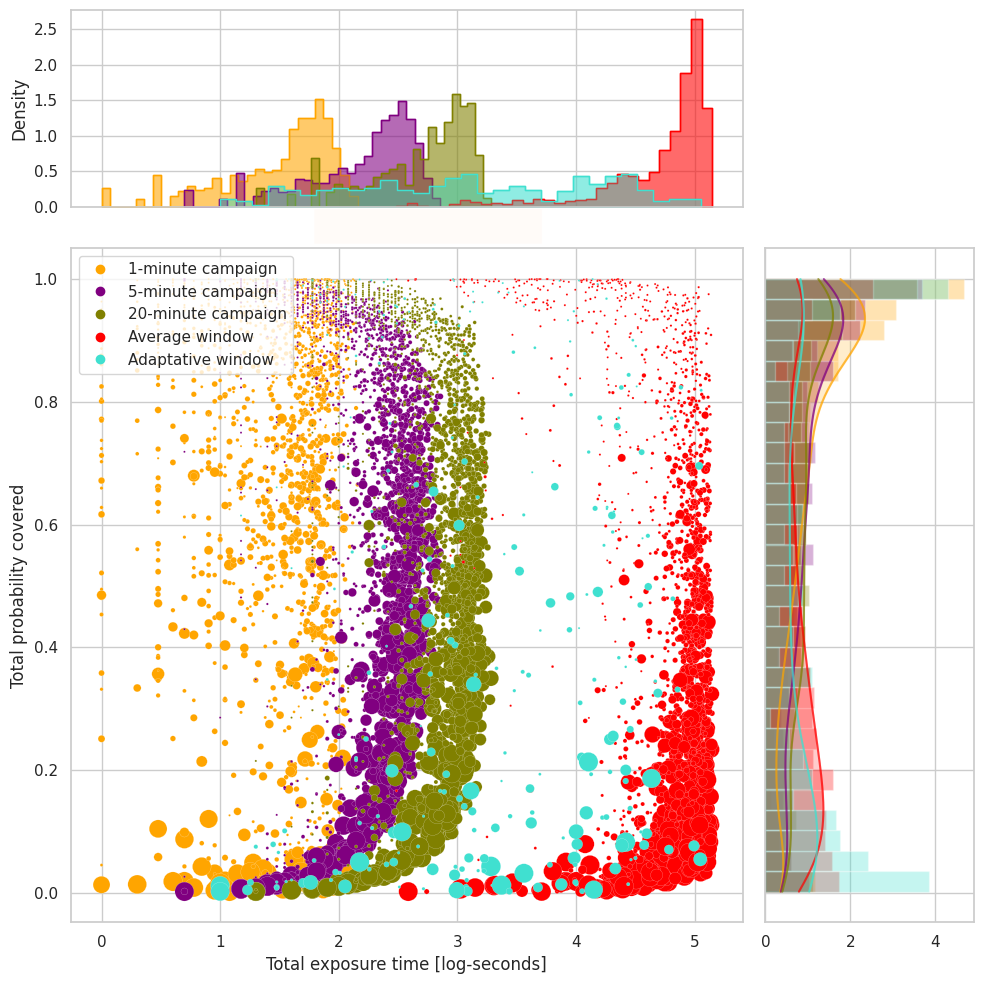}
    \caption{Total GW location uncertainty (named \textit{probability} for simplicity) covered in terms of the total exposure time of the for the 5 configurations explored in this work. The size of the markers corresponds to the size of the GW uncertainty regions, depicted for qualitative considerations.}
    \label{fig:scheduling-summary}
\end{figure}

The results of the five configurations are shown in Figure \ref{fig:scheduling-summary}. The total probability covered versus the total exposure time is presented, with their distributions projected for the horizontal and vertical dimensions. We identify the characteristics of the various scheduling simulations. In purely scheduling terms, the fixed duration campaigns, the 1', 5' and 20'-minute campaign, show how we can cover large C.R. with a very limited number of observations. Then, the average observation campaign scenario is not able to cover a large GW localization  uncertainty region. This effect is driven by the off-axis GRB sub-sample which require long exposure times at low latencies, being sub-optimal in realistic scenarios where no detailed information on the GRB emission is available beforehand. Similarly, a relatively low number of pointings are scheduled in the adaptive campaign, as many dimmer, off-axis GRBs do not meet the detectability requirement for scheduling.

\begin{figure}[h!]
    \includegraphics[width=0.98\linewidth]{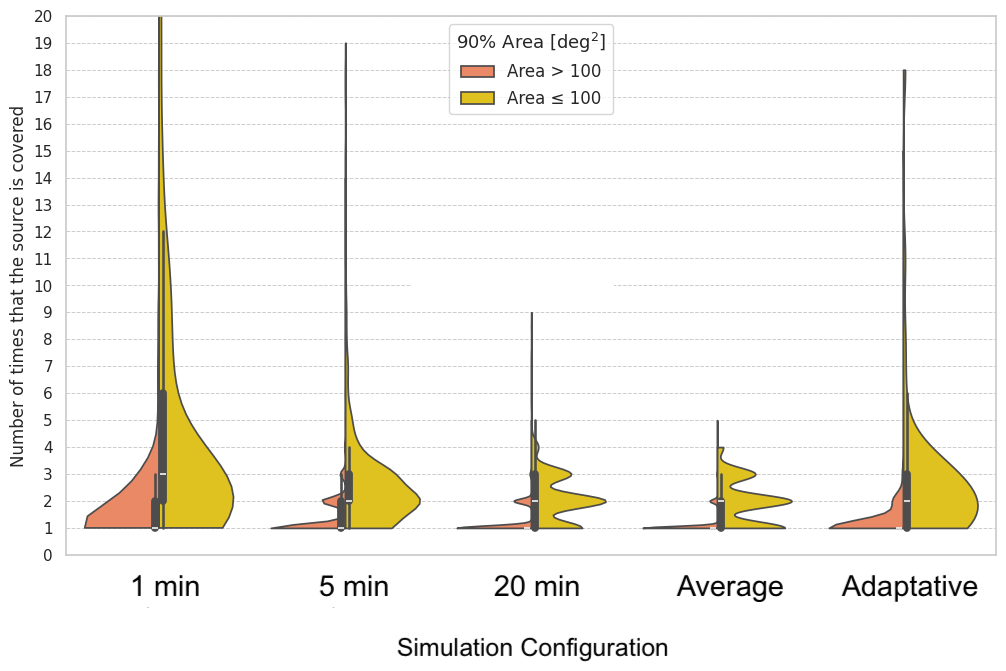}
    \caption{Distribution of the number of times that the GW-GRB source location is covered during the follow-up observation campaign of CTAO, for the different strategies presented in this work.}
    \label{fig:n-covered}
\end{figure}

Tiled observations can also contain a certain degree of sky overlap, enabling the coverage of some sky regions multiple times. Figure \ref{fig:n-covered} shows the number of times that the coordinates of the injected GW-GRB simulation are covered in each approach. In the absence of information about the energetics and temporal evolution of the GW-GRB source, short individual observations enable repeated coverage of the source location. In the most extreme case — 1-minute campaigns — the source can be revisited up to several tens of times, particularly for well-localized events. In Figure \ref{fig:n-covered}, we observe that the tested adaptive campaigns, including average- and variable-window campaigns do also produce such scenarios, i.e., the selected windows coincide with the fixed-duration, short window results. These cases correspond to well-localized events in combination with favorable energetics.

\subsection{Detection rates for follow-up strategies}
\label{sub:followup-detection-rates}

\begin{figure*}[tb]
    \centering

    \begin{subfigure}{0.45\textwidth}
        \includegraphics[width=\linewidth]{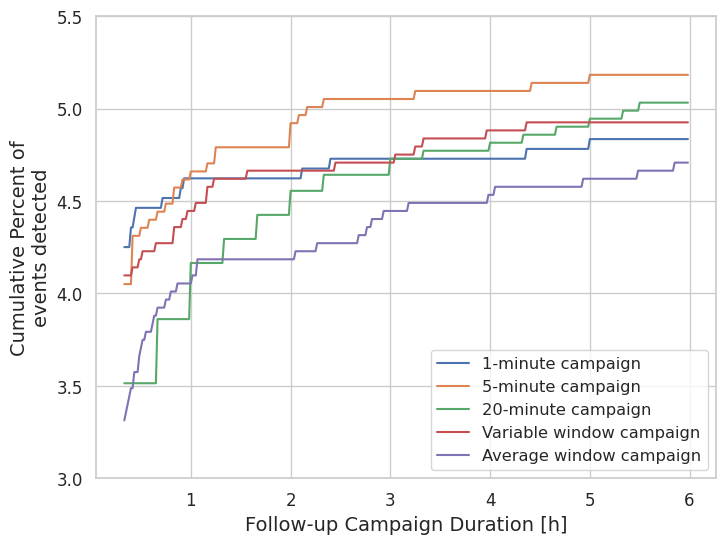}
        \caption{All events}
        \label{fig:cum-detections-followup-general}
    \end{subfigure}
    \hfill
    \begin{subfigure}{0.45\textwidth}
        \includegraphics[width=\linewidth]{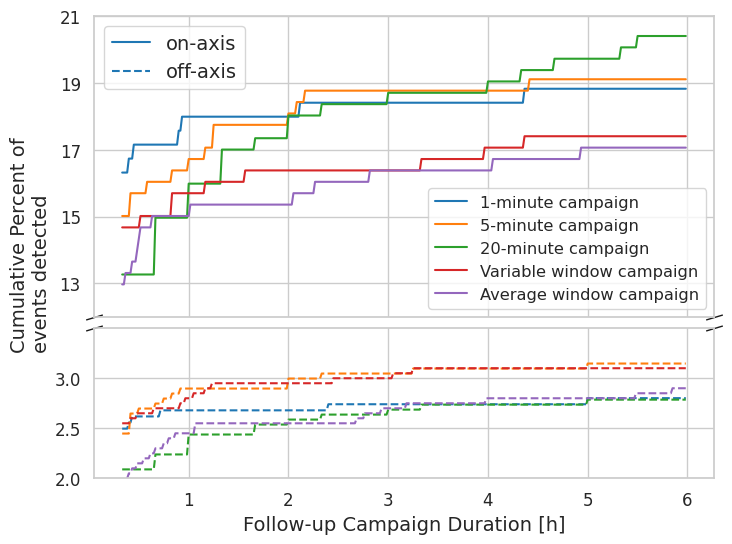}
        \caption{On- and off-axis events}
        \label{fig:cum-detections-followup-axis}
    \end{subfigure}

    \caption{Evolution of detection rates with increasing duration of follow-up campaign starting from the time when the first pointing is possible. An event can reach a $5\sigma$ detection in one single pointing, or when the cumulative significance of multiple correct pointings reaches the same threshold. The left panel compares the cumulative percentage of detected events for five different followup strategies, and shows that overall, after 3-4~hrs, the effectiveness of the variable, 1- and 5-minute campaigns stagnate. The right panel shows the on- and off-axis detection rates for the same strategies, each normalized to the number of events in their respective sample. Notably, for off-axis events, little seemed to be gained after following any strategy beyond $\sim1$~hr.}
    \label{fig:cum-detections-followup}
\end{figure*}

Just because a source location is covered quickly or with many repeated tilings does not imply that the event will be detected.  In order to fully evaluate the performance of the various scheduling algorithms, the tiling strategies must be folded with the detectability of each source. The results on the number of detected events versus follow-up campaign duration are shown in Fig.~\ref{fig:cum-detections-followup}, where the number of detectable on-axis and off-axis GRBs is shown. Across all strategies, increasing the duration of follow-up campaigns beyond $\sim3-4$~hrs provides rapidly diminishing benefits in terms of the number of events detected. When looking at off-axis GRBs only, this threshold decreases to $\sim1$~hr. This indicates that a follow-up campaign duration of $\sim4$~hr would be optimal to balance detection probability with telescope time used.

Table~\ref{tab:campaign-comparison} highlights summary statistics comparing each of the proposed campaign strategies, indicating that all three fixed-window campaigns manage to cover between 60-70\% of all followed-up events. With a fixed 4-hr follow-up, $4.5-5.1$\% of events are consistently detected regardless of the chosen strategy. However, the shorter 1- and 5-minute campaigns tend to produce higher detection significances, as the faster tiling provides a greater chance to reach the source position sooner or even cover it multiple times.

\begin{deluxetable*}{lRRRRR}
\tablecaption{Summary statistics for each of the five follow-up campaign scenarios. Coverage refers to whether the source position was successfully covered at least once by the scheduler. Source duration refers to the amount of time observing the correct source position. Lastly, in scenarios with a robust Real-Time Analysis (RTA) system, the scheduler can stop its scanning campaign and observe the correct source position as soon as a $5\sigma$ detection has been reached. \label{tab:campaign-comparison}}
\tablehead{
  \colhead{} &
  \colhead{1-minute} &
  \colhead{5-minute} &
  \colhead{20-minute} &
  \colhead{Variable} &
  \colhead{Average}
}
\startdata
\cutinhead{Percent (\%)}
Percent Covered                            & 70.0 & 65.9 & 62.9 & 6.7 & 44.7 \\
Percent Detected [4hr campaign]            &  4.7 & 5.1 & 4.8 & 4.9 & 4.5 \\
\cutinhead{Time (min)}
Mean time until source position reached    & 136.3 & 186.5 & 260.6 & 158.0 & 217.1 \\
Mean source duration [no RTA, 4hr campaign]&   0.6 & 1.7 & 4.4 & 0.3 & 2.5 \\
Mean source duration [RTA, 4hr campaign]   & 204.7 & 193.0 & 184.1 & 204.6 & 197.5 \\
\cutinhead{Significance ($\sigma$)}
Median Sig.\ [no RTA, 4hr campaign]        & 29.2 & 22.5 & 8.3 & 11.2 & 19.9 \\
Median Sig.\ [RTA, 4hr campaign]           & 91.4 & 49.3 & 43.1 & 83.7 & 58.5 \\
\enddata
\end{deluxetable*}

\subsection{The role of the Science Alert Generation}

We can further evaluate the performance of each strategy by considering how the internal, self-triggering alerts are produced by the SAG. Specifically, we focus on the Real Time Analysis (RTA). The RTA pipeline reconstructs the gamma-ray direction and energy from signals detected in the telescope cameras. It then generates and analyzes sky images in real time, concurrently assessing the detectability of observed sources over time. For dedicated GW-GRB campaigns, an RTA system strategically optimized for rapid response is essential. Such a system would identify signal hotspots within the telescope's field of view, allowing for an immediate cessation of the tiling schedule and a focused reallocation of observational resources to the detected hotspot.

\begin{figure}[tb]
    \centering
    \includegraphics[width=0.45\textwidth]{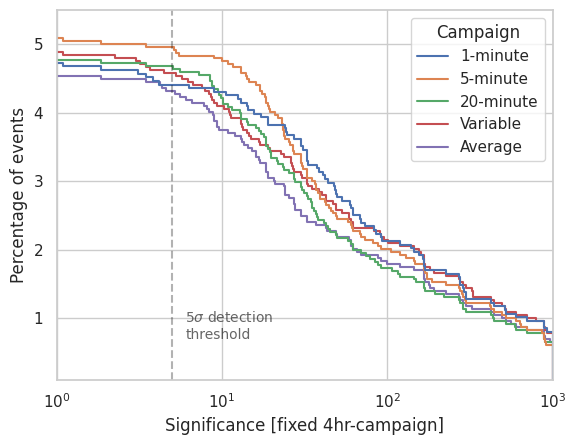}
    \caption{Distribution of the accumulated significance, displaying the detection and characterization capabilities of CTAO for the GW-GRB sample when the SAG system is involved, given a fixed 4~hr follow-up campaign with each strategy. In all cases, a SAG-RTA system was simulated, which stops the scheduler once a $5\sigma$ hotspot is detected in the FoV, and self-triggers observations of the source for the remaining duration of the campaign.}
    \label{fig:significance-rta}
\end{figure}

Figure~\ref{fig:significance-rta} shows the distribution of accumulated significance for a fixed-duration campaign with RTA, indicating that with RTA, $\sim4\%$ ($\sim2\%$) of total events can be detected at a $\ge10\sigma$ ($\ge100\sigma$), regardless of the chosen strategy.

Upon revisiting Table~\ref{tab:campaign-comparison}, we can also note the crucial role of the RTA in increasing the statistics of GRB detections beyond the \(5\sigma\) threshold. Fixing the remaining campaign time at the correct position doubles the detection significance on average and increases it by more than tenfold in some cases. In addition, the amount of observation time used to directly observe the source position increases by 2~-~3 orders of magnitude. The importance of RTA cannot be overstated, as the extra time directly leads to better resolved spectra, the possibility for detailed light curves, and higher statistics at energies near the CTAO threshold.

\section{Discussion on joint rates estimates} \label{sec:discussion}

The feasibility of joint detections of GW and sGRBs by the LVK and CTAO is estimated to be around 4 - 5\%. This joint detection rate is contingent upon several hypotheses, particularly the BNS merger rate and the successful jet fraction of GRBs, both parameters carrying considerable uncertainty.

\subsection{Impact of the uncertainties on CBC populations}

The joint detection rates of GW-GRB events depends on the BNS merger rate. \cite{2022ApJ...924...54P} referred to the published observational rate density estimates from the third LVK GW transient catalog (GWTC-3): $R^{GWTC3}_{BNS}= 10-1700  \mathrm{~Gpc}^{-3}\mathrm{yr}^{-1}$, as 90\% credible interval obtained by combining the rates estimated from different population models \citep{KAGRA:2021duu}. The most recent value, constrained including results from the first part of the fourth LVK observing run (O4a) and reported in the fourth LVK transient catalog (GWTC-4) is instead $R^{GWTC4}_{BNS}= 7.6-250  \mathrm{~Gpc}^{-3}\mathrm{yr}^{-1}$; also in this case, the numbers refer to the 90\% credible interval obtained by combining the rates estimated from different population models \citep{2025arXiv250818083T}. In both \cite{KAGRA:2021duu} and \cite{2025arXiv250818083T}, the impact of the population model selection and systematics are outlined, which yields to differences in the predicted BNS rates of various orders of magnitude. 
If we consider the number of BNS in \cite{2022ApJ...924...54P}, used in this work, we expect to observe during observing run O5 $190^{+410}_{-130}$ BNS mergers per year, based on recent updates of the expected duration and sensitivity of the O5 observing run. The BNS simulations used in this work are compared with those provided by the International Gravitational Wave Network (IGWN), used in the latest released BNS rates using GWTC-3 (\citealp{kiendrebeogo2023updated}). In this later work, the considerations on the populations that yield to compact binary coalescences (CBC) include a continuous mass distribution across the various CBC sub-populations (i.e. the three populations from the combinations of pairs of neutron stars and black holes), instead of a truncated gaussian distribution per sub-population. The annual number of detected BNS during O5 doubles (\citealp[180$^{+220}_{-100}$]{kiendrebeogo2023updated}, \citealp[86$^{+171}_{-9}$]{2022ApJ...924...54P}). More recently, \cite{2025A&A...702A..54S} presented the detection probability of BNS in O5 with a simplified approach that is independent of the uncertain
mass distribution of the merging binaries, and that uses only the
information on the past number of detections, combined with an
estimate of the ratio between the sensitivity of the target run with
respect to the previous runs; based on this study, the number of BNS we expect to observe in O5 is 28$^{+44}_{-21}$.  Thus, we conclude that the uncertainties regarding future BNS detections, which include the LVK network sensitivity and the CBC populations, are an important part of the \textit{systematic} errors affecting our detection prospects. Regarding the localization uncertainty regions, the median 90$\%$ C.R. source localization obtained in \cite{2022ApJ...924...54P} and \cite{kiendrebeogo2023updated} barely changes, from which we conclude that the findings on observation strategies presented in this work will remain unaffected. Future GW detections will lead to tighter constraints on the BNS merger rate, thereby enhancing the precision of the prospects outlined in this work.

\subsection{Impact of the uncertainties on very-high-energy gamma rays from GRBs}

All scenarios considered in this work assume successful VHE emission resulting from beamed afterglow radiation characteristic of jetted GRBs.  Yet, the fraction of BNS mergers producing jetted GRBs remains uncertain. Recent studies estimate that between 20\% and 50\% of the entire BNS population may be able to produce a jet \citep{nikhil_sgrb_2022,salafia_structure_2022}.

In addition, we do not expect this scenario to apply universally. BNS mergers may instead produce choked jets—jets that fail to successfully break out—such as was initially proposed for GW170817 during the first week after the merger, before Very Long Baseline Interferometry (VLBI) observations confirmed the presence of a launched jet. Looking ahead, gamma-ray observations will play a critical role in constraining emission models, thereby refining our understanding of jet formation and geometry in these events. The LST-1 observations on the BOAT GRB221009 have highlighted the capability of gamma-ray measurement to place stringent constraints on jet models, including from late times observations \cite{abe2025grb} .  

\subsection{The role of the observing strategies in IACTs} 

Observation strategies are designed to align with the specific goals of each campaign and can be guided by either prior hypotheses about the source or by agnostic, data-driven considerations. In this work, we explore both approaches. For poorly localized, transient events with low occurrence rates, the chosen strategy is particularly critical, although it is often complicated by limited knowledge of individual GRB sources and the GRB population. Our primary objective is to strike a balance between search sensitivity and sky coverage within a fixed maximum observing window. 

In addition, there is the possibility that telescopes may split into multiple sub-arrays that scan different regions of the sky in parallel. As shown in \citet{seglar2024cross}, this multi-telescope strategy (or divergent pointing mode) is especially powerful in the case of large telescope arrays like CTAO, as the coverage at time $t^N_{cov}$ scales as $t^N_{cov}\sim 1/N\cdot t_{1}$ ($N$: number of telescopes, $t_{1}$: time to cover a region by a single telescope),  modulated by the effective FoV of the subset of telescopes selected. This improvement comes at the expense of decreasing the sensitivity of the observation given smaller arrays, although the largest improvement in sensitivity due to strong background rejection arises in the transition from single-telescope to stereoscopic observations. While modifying the scheduling workflow for the N-observatory case is not in the scope of this paper, we acknowledge that sub-array strategies will be key for effective follow-up campaigns of poorly localized transient events. 

Finally, we highlight that galaxy catalogs were not incorporated into the observational strategy discussed herein. These catalogs are standard tools in gravitational-wave follow-up campaigns, crucial for optimizing sky coverage by precisely identifying probable host galaxies, a methodology successfully demonstrated in the case of GW170817 \citep{2017ApJ...848L..12A}. While a comprehensive evaluation of the benefits of integrating galaxy catalogs is beyond the purview of this study, we nevertheless acknowledge their considerable potential to augment detection efficiency. This potential is intrinsically linked to ongoing improvements in the completeness of galaxy catalogs, such as the GLADE+ catalog \citep{2022MNRAS.514.1403D}, the DESI Legacy Imaging Surveys \citep{2019AJ....157..168D} or the catalogs that will be produced with the Vera Rubin Observatory\footnote{\url{https://www.lsst.org/about/dm/data-products}},  which promise deeper and more comprehensive coverage of potential host galaxies. 

\section{Summary} \label{sec:conclusion}

In this work, we have explored the prospects of joint detection of GW by LVK during the Observing Run O5, and the sGRB detections by CTAO, for which the emission has been estimated from phenomenological descriptions and the latest simulated CTAO instrument response functions (Alpha configuration). 
We evaluated the role of observation strategies and how these can modulate the chance of detecting EM counterparts to GWs, especially in poorly localized events. We highlighted the role of real-time analysis in enabling the self-triggering of observations, improving the significance of detections, which in turn will enable an in-depth study of astrophysical sources, such as the sGRBs evaluated here. The observational and macro-physical parameters used in our study can be constrained in the case of non-detection at VHE energies. 

The full simulation chain of short GRBs associated with GW events, referred to as GW-GRBs, shows that approximately 18\% of the GW-GRB sample emits VHE gamma-ray radiation detectable by CTAO. This fraction is largely driven by off-axis events, which dominate the GW-GRB population, with on-axis events being detectable in a much higher fraction (see section \ref{sec:obs-strategy}). When detectability is combined with effective observability, accounting for realistic scheduling and observatory visibility, follow-up, and tiling strategies, a set of useful prescriptions for future CTAO observational planning emerges. Notably, follow-up campaigns in regions of large GW localization uncertainty become significantly less effective beyond 1–4 hours after the event, with minimal gains in detected events for longer observation times. 

Our analysis reveals that observation strategies based on either fixed or variable tile durations do not present significant variations, with the spread in detectable event percentages across strategies remaining within 25\%. Intriguingly, the highest number of detected sources is achieved with 5-minute observations, reaching a 5.1\% detection rate with a fixed, 4~hr campaign. Although shorter, 1-minute exposure times facilitate faster source acquisition, they exhibit a distinct performance difference compared to 20-minute observations, consistent with theoretical predictions. Therefore, we conclude that a 5-minute strategy represents an effective trade-off, balancing the requirements for exposure time and spatial coverage necessary for source detection.

The implementation of real-time analysis is crucial for integrating signals from significant hot spots, thereby increasing the scientific information that can be extracted from each event. Similarly, we can infer that longer observation times for well-localized events can provide valuable constraints, even in the absence of a detection.

The key macrophysical parameters influencing detectability are the jet opening angle and the viewing angle. In this regard, obtaining even rough estimates of the viewing angle in GW alerts would enable CTAO to prioritize the most promising events, maximizing detection prospects and also constraining the physical parameters of the GW-GRB system.

We note that the strategies discussed here do not incorporate other potentially useful information, such as the distribution of galaxies within the GW uncertainty region—an element that was crucial in the localization of GRB 170817. This highlights that there is still room for improvement in the observational strategies, which will be explored in a future work.

We have assessed how the results presented in this work rely on the various assumptions made concerning observations, astrophysical populations and rates, and gamma-ray emission, so that changes on these can be easily incorporated into the predicted fractions of detectable events.

The full simulation chain described here, developed within the CTAO Consortium, will be employed in forthcoming studies to further characterize the physical conditions that give rise to VHE emission and to assess the impact of CTAO observations in constraining the TeV GW-GRB population and the environment of BNS mergers. This work can also be extended to explore other populations, such as neutron star–black hole (NSBH) mergers, and to consider future gravitational-wave interferometers, including the Einstein Telescope. For these next-generation instruments, a careful definition of selection criteria will be crucial to identify GW events of potential interest amid the large number of detections expected.

\begin{acknowledgments}
{\bf CTAO Consortium Acknowledgments }

We gratefully acknowledge financial support from the following agencies and organizations:

\bigskip

State Committee of Science of Armenia, Armenia;
The Australian Research Council, Astronomy Australia Ltd, The University of Adelaide, Australian National University, Monash University, The University of New South Wales, The University of Sydney, Western Sydney University, Australia; Federal Ministry of Education, Science and Research, and Innsbruck University, Austria;
Conselho Nacional de Desenvolvimento Cient\'{\i}fico e Tecnol\'{o}gico (CNPq), Funda\c{c}\~{a}o de Amparo \`{a} Pesquisa do Estado do Rio de Janeiro (FAPERJ), Funda\c{c}\~{a}o de Amparo \`{a} Pesquisa do Estado de S\~{a}o Paulo (FAPESP), Funda\c{c}\~{a}o de Apoio \`{a} Ci\^encia, Tecnologia e Inova\c{c}\~{a}o do Paran\'a - Funda\c{c}\~{a}o Arauc\'aria, Ministry of Science, Technology, Innovations and Communications (MCTIC), Brasil;
Ministry of Education and Science, National RI Roadmap Project DO1-153/28.08.2018, Bulgaria; 
The Natural Sciences and Engineering Research Council of Canada and the Canadian Space Agency, Canada; 
ANID PIA/APOYO AFB230003, ANID-Chile Basal grant FB 210003, N\'ucleo Milenio TITANs (NCN19-058), FONDECYT-Chile grants 1201582, 1210131, 1230345, and 1240904; 
Croatian Science Foundation, Rudjer Boskovic Institute, University of Osijek, University of Rijeka, University of Split, Faculty of Electrical Engineering, Mechanical Engineering and Naval Architecture, University of Zagreb, Faculty of Electrical Engineering and Computing, Croatia;
Ministry of Education, Youth and Sports, MEYS  LM2018105, LM2023047, EU/MEYS 
CZ.02.1.01/0.0/0.0/16\_013/0001403,\\ CZ.02.1.01/0.0/0.0/18\_046/0016007, \\ CZ.02.1.01/0.0/0.0/16\_019/0000754, \\ and CZ.02.01.01/00/22\_008/0004632,  Czech Republic; 
Academy of Finland (grant nr.317636 and 320045), Finland;
Ministry of Higher Education and Research, CNRS-INSU and CNRS-IN2P3, CEA-Irfu, ANR, Regional Council Ile de France, Labex ENIGMASS, OCEVU, OSUG2020 and P2IO, France; 
The German Bundesministerium f\"ur Forschung, Technologie und Raumfahrt (BMFTR), the Max Planck Society, the Deutsche Forschungsgemeinschaft (DFG, with Collaborative Research Centre 1491), and the Helmholtz Association, Germany; Department of Atomic Energy, Department of Science and Technology, India; Istituto Nazionale di Astrofisica (INAF), Istituto Nazionale di Fisica Nucleare (INFN), MIUR, Istituto Nazionale di Astrofisica (INAF-OABRERA) Grant Fondazione Cariplo/Regione Lombardia ID 2014-1980/RST\_ERC, Italy; 
ICRR, University of Tokyo, JSPS, MEXT, Japan; 
Netherlands Research School for Astronomy (NOVA), Netherlands Organization for Scientific Research (NWO), Netherlands; 
University of Oslo, Norway; 
Ministry of Science and Higher Education, DIR/WK/2017/12, the National Centre for Research and Development and the National Science Centre, UMO-2016/22/M/ST9/00583, Poland; 
Slovenian Research and Innovation Agency, grants P1-0031, I0-E018, J1-60014, Slovenia; 
South African Department of Science and Technology and National Research Foundation through the South African Gamma-Ray Astronomy Programme, South Africa; 
The Spanish groups acknowledge funds from ``ERDF A way of making Europe", the Spanish Ministry of Science, Innovation and Universities, and the Spanish Research State Agency (AEI) via MCIN/AEI/10.13039/501100011033, grants CNS2023-144504, PDC2023-145839-I00, PID2022-137810NB-C22, PID2022-139117NB-C41/C42/C43/C44, PID2022-136828NB-C41/C42, PID2022-138172NB-C41/C42/C43, PID2021-124581OB-I00, PID2021-125331NB-I00, and budget lines 28.06.000X.411.01 and 28.06.000X.711.04 of PGE 2023, 2024 and 2025; the ``Centro de Excelencia Severo Ochoa" program through grants no. CEX2019-000920-S, CEX2020-001007-S, CEX2021-001131-S; the ``Unidad de Excelencia Mar\'ia de Maeztu" program through grants no. CEX2019-000918-M and CEX2020-001058-M; the ``Ram\'on y Cajal" program through grant RYC2021-032991-I; and the ``Juan de la Cierva" program through grants no. JDC2022-048916-I and JDC2022-049705-I. They also acknowledge the projects with refs. PR47/21 TAU and TEC-2024/TEC-182, both funded by the Comunidad de Madrid regional government. Funds were also granted by the ``Consejer\'ia de Universidad, Investigaci\'on e Innovaci\'on" of the regional government of Andaluc\'ia (Refs.~AST22\_00001\_9 and AST22\_0001\_16) and ``Plan Andaluz de Investigaci\'on, Desarrollo e Innovaci\'on" (Ref. FQM-322); and by the ``Programa Operativo de Crecimiento Inteligente" FEDER 2014-2020 (Refs.~ESFRI-2017-IAC-12 and ESFRI-2020-01-IAC-12) and Spanish Ministry of Science, Innovation and Universities, 15\% co-financed by ``Consejer\'ia de Econom\'ia, Industria, Comercio y Conocimiento" of the Gobierno de Canarias regional government. The Generalitat de Catalunya regional government is also gratefully acknowledged via its ``CERCA" program and grants 2021SGR00426 and 2021SGR00679. Spanish groups were also kindly supported by European Union funds via the Horizon Europe Research and innovation programme under Grant Agreement no. 101131928; NextGenerationEU, grants no. PRTR-C17.I1, CT19/23-INVM-109, and the ‘MicroStars’ ERC, ref. 101076533. This research used computing and storage resources provided by the Port d'Informaci\'o Cient\'ifica (PIC) data center;
Swedish Research Council, Royal Physiographic Society of Lund, Royal Swedish Academy of Sciences, The Swedish National Infrastructure for Computing (SNIC) at Lunarc (Lund), Sweden; 
State Secretariat for Education, Research and Innovation (SERI) and Swiss National Science Foundation (SNSF), Switzerland; The National Research Foundation of Ukraine (project 2023.03/0149 and 2023.05/0024), Ukraine;
Durham University, Leverhulme Trust, Liverpool University, University of Leicester, University of Oxford, Royal Society, Science and Technology Facilities Council, UK; 
U.S. National Science Foundation, U.S. Department of Energy, Argonne National Laboratory, Barnard College, University of California, University of Chicago, Columbia University, Georgia Institute of Technology, Institute for Nuclear and Particle Astrophysics (INPAC-MRPI program), Iowa State University, the Smithsonian Institution, V.V.D. is funded by NSF grant AST-1911061, Washington University McDonnell Center for the Space Sciences, The University of Wisconsin and the Wisconsin Alumni Research Foundation, USA.
The research leading to these results has received funding from the European Union's Seventh Framework Programme (FP7/2007-2013) under grant agreements No~262053 and No~317446.
This project is receiving funding from the European Union's Horizon 2020 research and innovation programs under agreement No~676134.

\bigskip

F. Sch\"ussler acknowledges ANR (French National Research Agency) for its support of the project "Multi-messenger Observations of the Transient Sky (MOTS)" under grant no. ANR-22-CE31-0012. 

A. Colombo and A. Stamerra acknowledge partial financial support by Italian Research Center on High Performance Computing Big Data and Quantum Computing (ICSC), project funded by European Union - NextGenerationEU - and National Recovery and Resilience Plan (NRRP) - Mission 4 Component 2 within the activities of Spoke 2 (Fundamental Research and Space Economy), (CN 00000013 - CUP C53C22000350006) 

A. Stamerra acknowledges financial support from INAF through the “Ricerca Fondamentale 2024” on the GO/GTO program titled ‘Discovering the New TeV Frontier of Gravitational Wave Counterparts through Follow-up Observations with the MAGIC Telescopes'.

L. Nava acknowledges financial support from  the European Union-Next Generation EU, PRIN 2022 RFF M4C21.1 (202298J7KT - PEACE)

\end{acknowledgments}

\begin{contribution}
This publication is the result of a collaborative effort within the CTAO Consortium, developing an integrated simulation framework connecting gravitational-wave detections, short gamma-ray burst emission models, and CTAO response simulations to assess the observatory’s capability to detect very-high-energy counterparts of compact binary mergers.
The project was coordinated by A. Stamerra, who supervised its development, guided the scientific discussion, and oversaw the manuscript preparation. L. Nava led the simulation of the intrinsic and time-dependent spectra of short gamma-ray bursts and their VHE emission, providing the phenomenological assumptions used in the study. B. Patricelli implemented and adapted the simulations of gravitational-wave–detected binary neutron star systems and contributed to the subsequent CTAO observation simulations. J. Green led the development of the dedicated package \texttt{sensipy} for simulating CTAO observations, carried out the analysis, and produced associated results and figures. M. Seglar Arroyo implemented the observing strategies within \texttt{tilepy}, generated the tiled observations used to scan the gravitational-wave skymaps, and produced the figures. F. Sch\"ussler led the \texttt{tilepy} development and contributed to the scientific discussions. All the authors participated in the interpretation of the results, preparation of the manuscript, and editing. 

The rest of the authors have contributed in one or several ways to the development of the CTAO analysis tools and IRFs and participated in the review of the manuscript through internal discussions and approval processes coordinated by the Transients Working Group leaders and the CTAO Speakers and Publications Office (CTAO-SAPO).
The initial manuscript drafts were reviewed by E. Bissaldi, F. Longo and I. Agudo on behalf
of CTAO-SAPO.  All authors above have participated in the paper discussion and edition.

\end{contribution}

\software{sensipy \citep{green_sensipy_2026}
tilepy \citep{seglar2024cross}
}


\appendix

\section{Derivation of temporal decays of VHE lightcurves and X-ray and TeV luminosities}\label{appendix_temporal slopes}
To derive the temporal decay index $\beta_2$ of the VHE light curves during the deceleration phase, it is assumed that they follow the same distribution as the X-ray lightcurves of real short GRBs.
We therefore considered a sample of short GRBs observed by Swift/XRT with data sufficiently sampled to allow fitting of the unabsorbed X-ray light curve in the 0.3-10\,keV energy band using a (multiple broken) power law model. The decay index during the deceleration phase is then identified as the slope of the final power-law segment. This choice is made to exclude the initial steep decay and plateau phase for consistency with our model (see Sect.~\ref{sec:VHE_emission}), which considers only standard temporal decays produced by the jet deceleration in interactions with the circumburst medium.

\begin{figure}
    \centering
    \includegraphics[width=0.8\linewidth]{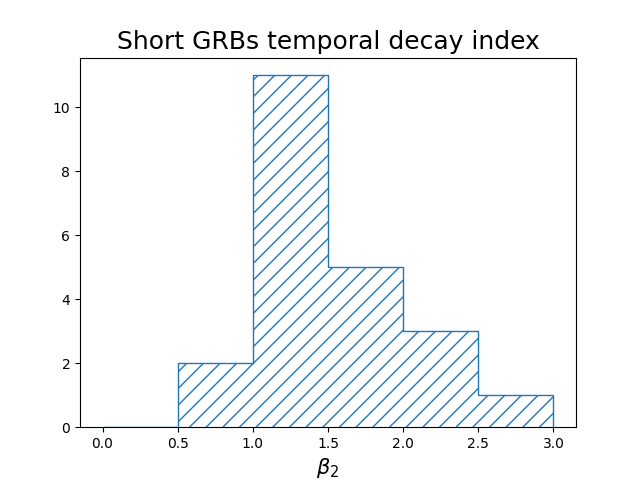}
    \caption{Distribution of the temporal decay indices $\beta_2$ ($F\propto t^{-\beta_2}$) of the X-ray afterglow of short GRBs.
    The temporal decay is obtained from a (multiple broken) power-law fit to the Swift/XRT unabsorbed light curve in the energy range 0.3-10\,keV (observer frame).}
    \label{fig:beta2_distribution}
\end{figure}

The resulting distribution is shown in Fig.~\ref{fig:beta2_distribution}: the mean value is 1.45 (median 1.37) and the standard deviation is 0.48.

\begin{figure}
    \centering
    \includegraphics[width=0.8\linewidth]{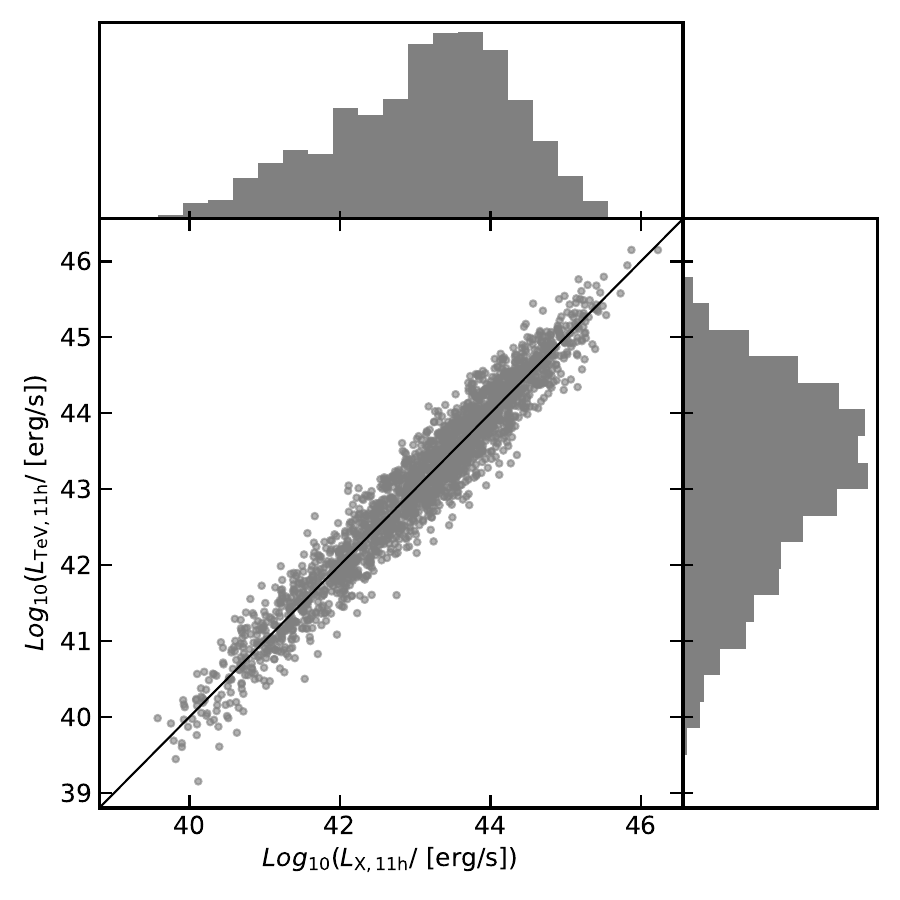}
    \caption{X-ray (0.3-10\,keV) and VHE (0.3-1\,TeV) simulated luminosities at 11\,h.}
    \label{fig:luminosities}
\end{figure}

The scatter plot for the inferred luminosities at 11\,h in X-ray and VHE energy range (see Sect.~\ref{sec:VHE_emission_lightcurves} of the main text for their derivation) are shown in Fig.~\ref{fig:luminosities}, together with their distributions.

\bibliography{references}{}
\bibliographystyle{aasjournalv7}



\end{document}

%% file: cta-authorlist.tex
\author[orcid=0000-0001-7250-3596,gname={Shotaro},sname={Abe}]{S.~Abe}
\affiliation{Institute for Cosmic Ray Research, University of Tokyo, 5-1-5, Kashiwa-no-ha, Kashiwa, Chiba 277-8582, Japan}
\email{shotaro@icrr.u-tokyo.ac.jp}

\author[orcid=0000-0001-8215-4377,gname={Jayant},sname={Abhir}]{J.~Abhir}
\affiliation{ETH Z\"urich, Institute for Particle Physics and Astrophysics, Otto-Stern-Weg 5, 8093 Z\"urich, Switzerland}
\email{jabhir@student.ethz.ch}

\author[orcid=0009-0005-5239-7905,gname={Abhishek},sname={Abhishek}]{A.~Abhishek}
\affiliation{INFN and Universit\`a degli Studi di Siena, Dipartimento di Scienze Fisiche, della Terra e dell'Ambiente (DSFTA), Sezione di Fisica, Via Roma 56, 53100 Siena, Italy}
\email{a.abhishek@student.unisi.it}

\author[orcid=0000-0002-6606-2816,gname={Fabio},sname={Acero}]{F.~Acero}
\affiliation{Universit\'e Paris-Saclay, Universit\'e Paris Cit\'e, CEA, CNRS, AIM, F-91191 Gif-sur-Yvette Cedex, France}
\affiliation{FSLAC IRL 2009, CNRS/IAC, La Laguna, Tenerife, Spain}
\email{fabio.acero@cea.fr}

\author[orcid=0000-0002-2028-9230,gname={Atreya},sname={Acharyya}]{A.~Acharyya}
\affiliation{University of Alabama, Tuscaloosa, Department of Physics and Astronomy, Gallalee Hall, Box 870324 Tuscaloosa, AL 35487-0324, USA}
\email{atreya@cp3.sdu.dk}

\author[orcid=0009-0000-5380-1109,gname={R\'emi},sname={Adam}]{R.~Adam}
\affiliation{Universit\'e C\^ote d'Azur, Observatoire de la C\^ote d'Azur, CNRS, Laboratoire Lagrange, France}
\affiliation{Laboratoire Leprince-Ringuet, CNRS/IN2P3, \'Ecole polytechnique, Institut Polytechnique de Paris, 91120 Palaiseau, France}
\email{remi.adam@oca.eu}

\author[orcid=0000-0001-8816-4920,gname={Arnau},sname={Aguasca-Cabot}]{A.~Aguasca-Cabot}
\affiliation{Departament de F{\'\i}sica Qu\`antica i Astrof{\'\i}sica, Institut de Ci\`encies del Cosmos, Universitat de Barcelona, IEEC-UB, Mart{\'\i} i Franqu\`es, 1, 08028, Barcelona, Spain}
\email{arnau.aguasca@fqa.ub.edu}

\author[orcid=0000-0002-3777-6182,gname={Iv\'an},sname={Agudo}]{I.~Agudo}
\affiliation{Instituto de Astrof{\'\i}sica de Andaluc{\'\i}a-CSIC, Glorieta de la Astronom{\'\i}a s/n, 18008, Granada, Spain}
\email{iagudo@iaa.es}

\author[orcid=0009-0006-6223-5018,gname={Irene},sname={Albanese}]{I.~Albanese}
\affiliation{INFN Sezione di Padova and Universit\`a degli Studi di Padova, Via Marzolo 8, 35131 Padova, Italy}
\email{irene.albanese@cta-consortium.org}

\author[gname={Jorge},sname={Alfaro}]{J.~Alfaro}
\affiliation{Pontificia Universidad Cat\'olica de Chile, Av. Libertador Bernardo O'Higgins 340, Santiago, Chile}
\email{jalfaro@fis.puc.cl}

% ...

{\renewcommand\thefootnote{}\footnotetext{\hskip-11pt* Corresponding author}}

\collaboration{all}{({CTAO Consortium})}

\collaboration{all}{{Other co-authors}}
\author[orcid=0000-0002-7439-4773,gname={Alberto},sname={Colombo}]{A. Colombo}
\affiliation{INAF - Osservatorio Astronomico di Roma, Via di Frascati 33, 00078, Monteporzio Catone, Italy}
\affiliation{Dep. of Physics, Sapienza, University of Roma, Piazzale A. Moro 5, 00185, Roma, Italy}
\email{alberto.colombo@inaf.it}